\documentstyle[12pt,epsfig]{article}
\newfont{\Bbb}{msbm10 scaled 1200}     
\newcommand{\mathbb}[1]{\mbox{\Bbb #1}}

\def\IR{{\mathbb R}}
\def\IZ{{\mathbb Z}}

\def\TL{\hfil$\displaystyle{##}$}
\def\TR{$\displaystyle{{}##}$\hfil}

\def\lbldef#1#2{\expandafter\gdef\csname #1\endcsname {#2}}
\def\eqn#1#2{\lbldef{#1}{(\ref{#1})}%
\begin{equation} #2 \label{#1} \end{equation}}
\def\eqalign#1{\vcenter{\openup1\jot
    \halign{\strut\span\TL & \span\TR\cr #1 \cr
   }}}

\def\href#1#2{#2}  

\def\ct{\cos t}
\def\st{\sin t}
\def\chr{ \cosh \rho}
\def\shr{ \sinh \rho}
\def\sp{ \sin \phi}
\def\cp { \cos \phi}
 
\newcommand{\beq}{\begin{equation}}
\newcommand{\eeq}{\end{equation}}
\newcommand{\ber}{\begin{eqnarray}}
\newcommand{\eer}{\end{eqnarray}}

\newcommand{\beqar}{\begin{eqnarray}}

\newcommand{\eeqar}{\end{eqnarray}}

\begin{document}
\baselineskip=15.5pt
\pagestyle{plain}
\setcounter{page}{1}
\begin{titlepage}

\leftline{\tt hep-th/0001053}

\vskip -.8cm

\rightline{\small{\tt CALT-68-2245}}
\rightline{\small{\tt CITUSC/99-010}} 
\rightline{\small{\tt HUTP-99/A027}}
\rightline{\small{\tt LBNL-44375}}
\rightline{\small{\tt  UCB-PTH-99/48}}

\begin{center}

\vskip 1.7 cm

{\LARGE { Strings in $AdS_3$ and the $SL(2,R)$ WZW Model. }}
\vskip .5cm
{\LARGE {Part 1: The Spectrum}}

\vskip 1.5cm
{\large 
Juan Maldacena$^{*}$ and
Hirosi Ooguri$^{\dagger}$\footnote{On leave of absence
from the University of California, Berkeley.}}

\vskip 1.2cm

${}^*$ Lyman Laboratory of Physics,
Harvard University, Cambridge, MA  02138, USA

\medskip
${}^\dagger$ Caltech - USC Center for Theoretical Physics, 
Mail Code 452-48 
\\
California Institute of Technology, 
Pasadena, CA 91125, USA 

\vskip 0.5cm

{\tt   malda@pauli.harvard.edu, ooguri@theory.caltech.edu}

\vspace{1cm}

{\bf Abstract}
\end{center}

\noindent
In this paper we study the spectrum of bosonic string theory
on $AdS_3$. We study classical solutions of the $SL(2,R)$ WZW model,
including solutions for long strings with non-zero winding number.
We show that the model has a symmetry relating string
configurations with different winding numbers. 
We then study the Hilbert space of the WZW model, including 
all states related by the above symmetry. This leads to 
a precise description of long strings. We prove a no-ghost
theorem for all the representations that are involved and
 discuss the scattering of the long string.

\end{titlepage}

\newpage


\section{Introduction}
\label{intro}

In this paper we study the spectrum of critical
bosonic  string theory on $AdS_3 \times {\cal M}$ with
NS-NS backgrounds, where ${\cal M}$ is a compact space.
Understanding string theory on $AdS_3$ is interesting from the
point of view of the $AdS$/CFT correspondence since it enables 
us to study the correspondence beyond the gravity approximation. 
Another motivation is to understand string theory on a curved
space-{\underline{time}}, where the timelike component $g_{00}$ 
of the metric is non-trivial.

This involves understanding the $SL(2,R)$ WZW model. In this paper,
we always consider the case when the target space is 
the universal cover of the $SL(2,R)$ group manifold so that the
timelike direction is non-compact.  
The states of the WZW model form representations of
the  current algebras $\widehat{SL}(2,R)_L \times \widehat{SL}(2,R)_R$.
Once we know which representations of these algebras appear, 
we can find the physical states of a string in $AdS_3$ by imposing the 
Virasoro constraints on the representation spaces. The problem is to 
find the set of representations that one should consider. 
In WZW models for compact groups, the unitarity restricts the 
possible representations \cite{Gepner:1986wi}.
Representations of $\widehat{SL}(2,R)$, on the other hand, are not unitary
except for the trivial representation. Of course this is not a surprise; 
the physical requirement is that states should have non-negative norms
only after we impose the Virasoro constraints. 
Previous work on the subject 
\cite{Balog:1989jb,Petropoulos:1990fc,Mohammedi:1990dp,%
Bars:1991rb,Hwang:1991aq,%
Hwang:1992an,Hwang:1998tr,Evans:1998qu,Petropoulos:1999nc} typically 
considered representations with $L_0$ bounded
below and concluded that the physical spectrum does not contain
negative norm states if there is the restriction  $0<j<k/2$ on 
the $SL(2,R)$ spin $j$ of the representation; the spin 
of the $SL(2,R)$ is roughly the mass of the string state in $AdS_3$.

This restriction raises two puzzles. One is that it seems to imply an
upper bound on the mass of the string states in $AdS_3$ so that
the internal energy of the string could not be too high. For example,
if the compact space ${\cal M}$ has a nontrivial $1$-cycle, we find
that there is an upper bound on the winding number on the cycle.  
This restriction, which is independent of the string 
coupling, looks very arbitrary and raises doubts about the consistency 
of the theory. 
The second puzzle is that, on physical grounds, we expect that
the theory contains states corresponding
to the long strings of \cite{Maldacena:1998uz,Seiberg:1999xz}.
These are finite energy states where we have a long string stretched 
close to the boundary of $AdS_3$.   
These states are not found in any representation 
with $L_0$ bounded below. 
In this paper, we propose that the Hilbert space of the WZW model
includes a new type of representations, and we show that this
proposal resolves both the puzzles. In these new representations, 
$L_0$ is not bounded below. They are obtained by acting on
the standard representations by elements of the loop group that
are not continuously connected to the identity, through an operation
called spectral flow. 
These representations in the $SL(2,R)$ WZW model
have also been considered, with some minor variations, in 
\cite{Henningson:1991jc,Hwang:1992uk}. The authors of these
papers were motivated by finding a modular invariant partition 
function. They were, however, considering the case when the target
space is $SL(2,R)$ group manifold and not its universal cover. 

Throughout this paper, we consider $AdS_3$ in  global coordinates, which
do not have a coordinate horizon. In these coordinates,
the unitarity issue becomes clearer since strings cannot fall behind
any horizon. 
The interested reader could refer to 
\cite{Bars:1995cn,Bars:1996mf,Bars:1999ik} for studies 
involving $AdS_3$ in Poincare coordinates.
{}From the point of view of the $AdS$/CFT correspondence, it 
is the  spectrum of strings on $AdS_3$ in the global coordinates that
determines
the spectrum of conformal dimensions of operators in the boundary 
CFT, though in principle the same information could be extracted from 
the theory in Poincare coordinates. 

In order to completely settle the question of consistency of the 
$SL(2,R)$ WZW model, one needs to show that the OPE of two elements of
the set of representations
that we consider contains only elements of this set. We plan to
discuss this issue in our future publication.

\medskip

The organization of this paper is as follows. 
In section 2, we study classical solutions of the $SL(2,R)$ WZW model
and we show that the model has a spectral flow symmetry which relates
various solutions. In section 3, we do a semi-classical analysis
and have the first glimpse of what happens when we raises the internal 
excitation of the string beyond the upper bound implied by the restriction 
$j<k/2$. In 
section 4, we study the full quantum problem and we propose a set 
of representations that gives a spectrum for the model with the
correct semi-classical limits. In section 5, 
we briefly discuss scattering amplitudes involving 
the long strings. We conclude the paper with a summary
of our results in section 6. 
In appendix A, we extend the proof of the no-ghost theorem
for the representations we introduced in section 4. In appendix B, 
we study the one-loop partition function in $AdS_3$ with the
Lorentzian signature metric and
show how the sum over spectral flow reproduces the result
\cite{Gawedzki:1991yu} after taking an  Euclidean signature metric, 
up to contact terms in the modular parameters
of the worldsheet.

\section{Classical solutions}
\label{class}

We start by choosing a parameterization of the $SL(2,R)$ group element as 
\eqn{elem}{\eqalign{
 g  &= e^{ i u \sigma_2 } e^{ \rho \sigma_3} e^{i v \sigma_2} 
\cr &=
\left(\begin{array}{ll}
\ct \chr + \cp \shr &
\st \chr - \sp \shr \\
-\st \chr - \sp \shr  &
\ct \chr - \cp \shr  \\
\end{array} \right) .}}
Here $\sigma^i$ ($i=1,2,3$) are the Pauli
matrices\footnote{$\sigma_1 = \left(\begin{array}{ll} 
 0 & 1 \\ 1 & 0 \end{array}\right) $,  $\sigma_2 = \left(\begin{array}{ll} 
 0 & -i \\ i & 0 \end{array}\right) $ and
$\sigma_3 = \left(\begin{array}{ll} 
 1 & 0 \\ 0 & -1 \end{array}\right) $. },
and we set 
\eqn{defuv}{
u = {1 \over 2} ( t + \phi) ~,~~~~~~~~ 
v = {1\over 2}( t- \phi) .  
}
Another useful parameterization of $g$ is
\eqn{elem2}{ g = 
 \left(\begin{array}{ll}
 X_{-1} + X_1  & X_0 - X_2 \\ -X_0 -X_2 & X_{-1} - X_1 
\end{array} \right),}
with
\eqn{hyperboloid}{ X_{-1}^2 + X_0^2 - X_1^2 - X_2^2 = 1. }  
This parameterization shows that the $SL(2,R)$ group manifold
is a $3$-dimensional hyperboloid. The metric on $AdS_3$,
$$ ds^2 = - dX_{-1}^2 - dX_0^2 + dX_1^2 + dX_2^2, $$
is expressed in the global coordinates $(t,\phi,\rho)$
as
\eqn{metr}{
ds^2 = - \cosh^2\rho dt^2 + d\rho^2 + \sinh^2\rho d\phi^2.
}
 We will always work
on the universal cover of the hyperboloid (\ref{hyperboloid}), 
and $t$ is non-compact.

Our theory has the WZW action
\eqn{wzwac}{
S = { k \over 8 \pi \alpha'}
 \int d^2 \sigma {\rm Tr} \left( g^{-1} \partial g g^{-1} \partial g
\right)+ k \Gamma_{WZ}
}
The level $k$ is not quantized since $H^3$ vanishes for $SL(2,R)$. 
The semi-classical limit corresponds to large $k$. 
We define the right and left moving coordinates on the
worldsheet as, 
\eqn{sigmas}{
x^\pm = \tau \pm \sigma,}
where $\sigma$ is periodic with the period $2\pi$.
This action has a set of conserved right and left moving currents
\eqn{currents}{
J_R^a(x^+) = k {\rm Tr}\left( T^a\partial_+ g g^{-1} \right), ~~~~~
J_L^a(x^-) = k {\rm Tr}\left( T^{a*} g^{-1} \partial_- g \right)
}
where $T^a$ are a basis for the $SL(2,R)$ Lie algebra. It is convenient to 
take them as
$$T^3 = -{i\over 2} \sigma^2,~~
T^{\pm} = {1 \over 2}( \sigma^3 \pm i \sigma_1).$$
In terms of our parameterization, the currents are expressed as 
\eqn{curright}{\eqalign{
J^3_R =& k( \partial_+u + \cosh 2 \rho  \partial_+v)  \cr
J^{\pm}_R = & k  
 (\partial_+ \rho \pm i  \sinh 2 \rho \partial_+ v  )
 e^{\mp i2 u }~, }}
and 
\eqn{curleft}{\eqalign{
J^3_L =& k(  \partial_-v   + \cosh 2 \rho  \partial_- u) \cr
J^{\pm}_L = & k
 (\partial_- \rho \pm i  \sinh 2 \rho \partial_- u )
 e^{\mp i 2  v } ~.}}
The zero modes of $J^3_{R,L}$ are related to the energy $E$ and angular
momentum $\ell $ 
in $AdS_3$ as
\eqn{energymomentum}{
\eqalign{ & J_0^3 = \int_0^{2\pi} {dx^+ \over 2\pi} J^3_{R} 
= {1 \over 2} (E + \ell ) \cr
&\bar{J}_0^3 =  \int_0^{2\pi} {d x^- \over 2\pi} 
J^3_L = {1 \over 2}(E -\ell).}} 
The second Casimir of $SL(2,R)$ is 
\eqn{casi}{
c_2 = J^a J^a = {1 \over 2} \left( J^+ J^- + J^- J^+ \right)
 - (J^3)^2~.
}

The equations of motion derived from (\ref{wzwac}) is
$\partial_-(\partial_+ g g^{-1}) = 0$, namely that the currents, 
$J_R$ and $J_L$, are purely right or left moving as indicated. 
A general solution of the equations of motion for $SL(2,R)$ is 
the product of two group elements each of which depends only on 
$x^+$ or $x^-$ as
\eqn{genso}{
 g = g_+(x^+)g_-(x^-).}
Comparing \genso\ with \elem\ we can find the embedding of the worldsheet 
in $AdS_3$. 
The requirement that the string
is closed under $\sigma \rightarrow \sigma + 2\pi$
imposes the constraint,
\eqn{worldsheetperiodicity}{
g_+(x^+ + 2\pi) = g_+(x^+ ) M,~~~~ 
g_-(x^- - 2\pi) = M^{-1} g_-(x^-),}
with the same $M \in SL(2,R)$ for both $g_+$ and $g_-$. 
The monodromy matrix $M$ is only defined up to a conjugation
by $SL(2,R)$, 
and classical solutions of the WZW model are 
classified according to the conjugacy class of $M$.

For strings on $AdS_3 \times {\cal M}$, 
we should impose the Virasoro constraints
\eqn{classicalvirasoro}{
T^{total}_{++} 
 = T_{++}^{AdS} + T^{other}_{++} =0 } 
and similarly  
$T^{total}_{--} = 0$, where 
$$T_{++}^{AdS}  = { 1 \over k} J^a_R J^a_R$$ 
is the energy-momentum tensor for the $AdS_3$ 
part\footnote{In the quantum
theory,
we will have the same expression but with $k \to k-2$.} 
and $T^{other}_{++}$ represents the energy-momentum tensor
for the sigma-model on ${\cal M}$. 

\medskip

Let us analyze some simple classical solutions.

\subsection{Geodesics in AdS$_{\bf 3}$}

\begin{figure}[htb]
\begin{center}
\epsfxsize=5.0in\leavevmode\epsfbox{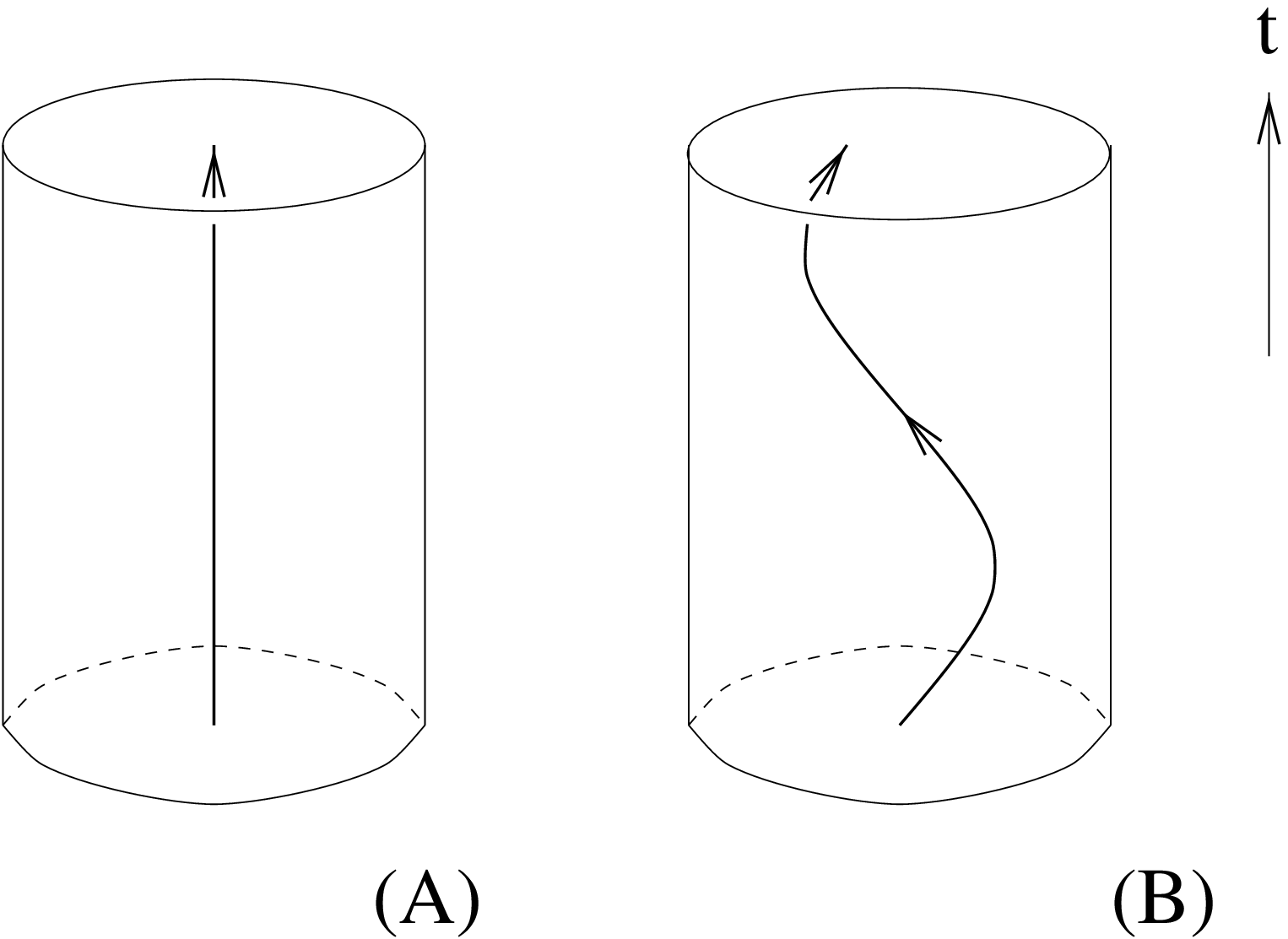}
\end{center}
\caption{Timelike geodesic; (A) a solution (\ref{timelikegeodesic})
with $U=V=1$, (B) a general geodesic is obtained by acting 
the $SL(2,R) \times SL(2,R)$ isometry on (A).}
\label{GF1}
\end{figure} 

\begin{figure}[htb]
\begin{center}
\epsfxsize=5.0in\leavevmode\epsfbox{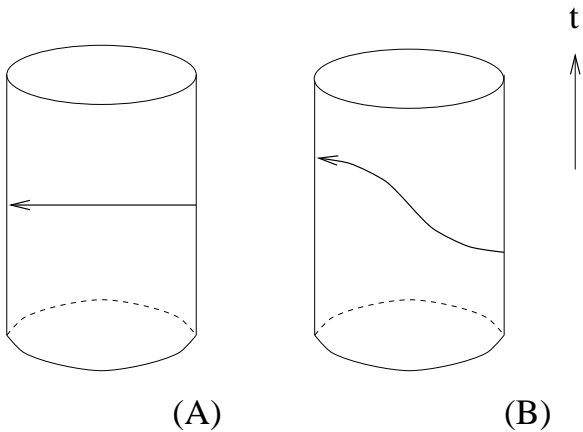}
\end{center}
\caption{Spacelike geodesic; (A) a solution (\ref{spacelikegeodesic})
with $U=V=1$, (B) a general geodesic is obtained by acting 
the $SL(2,R) \times SL(2,R)$ isometry on (A).}
\label{GF2}
\end{figure}

Consider a solution
\eqn{geodesicansatz}
{   g_+ = U e^{iv_+(x^+) \sigma_2}, 
~~~ g_- =  e^{iu_-(x^-) \sigma_2} V,}
where $U$ and $V$ are constant elements of $SL(2,R)$.  
The energy momentum tensor of this solution is 
\eqn{energytimelikegeodesic}
{T_{++}^{AdS} = - k (\partial_+ v_+)^2,~~ T_{--}^{AdS}
= - k (\partial_- u_-)^2 . }
Suppose we have some string excitation in the compact
part ${\cal M}$ of $AdS_3 \times {\cal M}$,  
and set $T_{\pm\pm}^{other} = h$ for some constant $h > 0$.
We may regard $h$ as a conformal weight of
the sigma-model on ${\cal M}$.
The Virasoro constraints $T_{\pm\pm}^{total}=0$ implies
$$ (\partial_+ v_+)^2 = (\partial_- u_-)^2 =  {h \over k}. $$
Thus we can set
$v_+ =  \alpha x^+/2$ and 
$u_- = \alpha x^-/2$
where $\alpha = \pm \sqrt{4h/k}$.
Substituting this in \genso, we obtain
\eqn{timelikegeodesic}
{ g = U \left(\begin{array}{ll}
 \cos(\alpha \tau) &
\sin(\alpha \tau) \\
-\sin(\alpha \tau)  &
 \cos(\alpha \tau) \\
\end{array} \right) V.}
Since the solution depends only on  $\tau$ 
and not on $\sigma$, we interpret that the string is 
collapsed to a point which flows along the trajectory 
in $AdS_3$ parameterized by $\tau$. See Figure \ref{GF1}. 
If $U=V=1$, the solution 
(\ref{timelikegeodesic}) represents a particle
sitting at the center of $AdS_3$, 
\eqn{simplegeodesic}
{t = \alpha \tau, ~~\rho=0.}
A more general solution (\ref{timelikegeodesic})
is given by acting the $SL(2,R) \times SL(2,R)$ isometry 
on (\ref{simplegeodesic}), and therefore it is
a timelike geodesic\footnote{In fact, any timelike geodesic
can be expressed in the form \timelikegeodesic . }
  in $AdS_3$. For this solution, 
the currents are given by
\eqn{currentformassiveparticle}
{ J^a_R T^a   = {k \over 2} \alpha U T^3 U^{-1},}
and similarly for $J_L$.  The monodromy matrix $M$ defined
by (\ref{worldsheetperiodicity}) is
$$ 
M = \left(\begin{array}{ll}
\cos(\alpha\pi) & \sin(\alpha\pi) \\
-\sin(\alpha\pi) &
\cos(\alpha\pi)  \\
\end{array} \right) $$
and belongs to the elliptic conjugacy
class $SL(2,R)$.

A solution corresponding to 
a spacelike geodesic is
\eqn{spacelikegeodesic}
{ g = U \left(\begin{array}{ll}
e^{\alpha \tau} & 0 \\
0 &
e^{-\alpha \tau }  \\
\end{array} \right) V,}
with $U, V \in SL(2,R)$. 
The energy-momentum tensor has a sign opposite of
(\ref{energytimelikegeodesic})
\eqn{energyspacelikegeodesic}
{ T^{AdS}_{\pm\pm} = {1 \over 4 } k \alpha^2.} 
If we choose $U = V = 1$, the solution is simply 
a straight line cutting the spacelike section $t=0$ 
of $AdS_3$ diagonally, 
\eqn{simplegeodesictwo}
{t=0,~~\rho e^{i\phi} = \alpha\tau , }
See Figure \ref{GF2} (A).
A general solution (\ref{spacelikegeodesic}) is given
from this by the action of the isometry, and therefore
is a spacelike geodesic. The currents for this solution are
\eqn{currentfortachyon}
{ J^a_R T^a  = {k \over 2} \alpha U T^1 U^{-1},}
and the monodromy matrix is
$$ 
M = \left(\begin{array}{ll}
e^{\alpha \pi} & 0 \\
0  &
e^{-\alpha \pi} \\
\end{array} \right), $$
which belongs to the hyperbolic conjugacy class of $SL(2,R)$. 

There is one more class of solutions whose monodromy matrices
are in the parabolic conjugacy class of $SL(2,R)$. They
correspond to null geodesics in $AdS_3$.

\subsection{Spectral flow and strings with winding numbers}

Given one classical solution $g=\tilde{g}_+\tilde{g}_-$,
we can generate new solutions by the following operation,
\eqn{classflow}{
g_+ = e^{i {1\over 2} w_R  x^+ \sigma_2} 
 \tilde g_+ ~~~~~~~~~ 
g_- = \tilde g_-  e^{i{1 \over 2} w_L  x^- \sigma_2}.
}
Comparing this with the parameterization (\ref{elem})
of $g=g_+ g_-$, we see that this operation amounts to
\eqn{classflowtwo}
{\eqalign{ & t \rightarrow
            t + {1 \over 2} (w_R + w_L) \tau 
      + {1 \over 2} (w_R - w_L) \sigma \cr
       & \phi 
  \rightarrow \phi + {1 \over 2} (w_R + w_L)\sigma +
       {1\over 2} (w_R- w_L) \tau .}}
The periodicity of the string worldsheet, under $\sigma \to \sigma + 2
\pi$,
 on the universal cover 
of $SL(2,R)$ requires\footnote{
If the target space is the single cover 
of $SL(2,R)$, $w_R$ and $w_L$ can be different.
In this case $(w_R - w_L)$ gives the winding number
along the closed timelike curve on $SL(2,R)$.}
$w_R = w_L = w$ for some integer $w$. 

One may regard \classflow\ as an action by an element of the
loop group $\widehat{SL}(2,R) \times \widehat{SL}(2,R)$ which is 
not continuously connected to the identity\footnote{
The loop group $\widehat{SL}(2,R)$ has such an element
since $\pi_1(SL(2,R))= \IZ$. Therefore, in the model
whose the target space is the single cover of $SL(2,R)$,
the full symmetry group of the model is the loop group
of $SL(2,R) \times SL(2,R)$ and its connected components are 
parametrized by $\IZ \times \IZ$. In this paper, we are
studying the model for the universal cover of $SL(2,R)$.
In this case, some of these elements do not act properly
on the field space, generating worldsheets which close only
modulo time translation. However the ones parametrized by the
diagonal $\IZ$ are still symmetry of the model. The diagonal 
$\IZ$ parameterizes the spectral flow operation performed 
simultaneously for both the left and right movers.}.
This particular symmetry of the theory will also be useful in our 
analysis of the Hilbert space. Here we see that it generates 
a new solution from an old
solution. Furthermore, the currents \curright\
 change in the following way
\eqn{changecur}{
J_R^3 = \tilde J^3_R + {k \over 2} w, 
~~~ J^\pm_R = \tilde J^\pm_R e^{ \mp i w
x^+}
}
and a similar expression for $J_L^a$. Or, in terms of the
Fourier modes,
\eqn{changecur2}
{ J_n^3 = \tilde{J}^3_n + {k \over 2} w \delta_{n,0},
~~~J_n^\pm = \tilde{J}_{n\mp w}^\pm.}
This means that the 
stress tensor will change to 
\eqn{changestr}{
T^{AdS}_{++} = \tilde  T^{AdS}_{++} - w\tilde J^3  - {k \over 4} w^2. 
}
In the CFT literature, this operation is known as the spectral flow. 

Let us study what happens if we act with this symmetry on the solutions 
corresponding to geodesics, \timelikegeodesic\ and \spacelikegeodesic. 
These solutions depend only on the worldsheet time coordinate
$\tau$, and the spectral flow \classflowtwo\ with $w = w_R = w_L$
introduces $\sigma$ dependence as
\eqn{rotatesolution}{
\eqalign{ t & = t_0(\tau) + w \tau \cr
          \rho & = \rho_0(\tau) \cr
           \phi & = \phi_0(\tau) + w \sigma .}}
Here $(t_0, \rho_0, \phi_0)$ represents the original geodesic solution. 
So what the spectral flow does is to stretch the geodesic
solution in the $t$-direction (by adding $w\tau$) and
rotates it around $w$-times around the center $\rho=0$ 
of $AdS_3$ (by adding $w\sigma$).
It is clear that the resulting solution describes
a circular string, winding $w$-times around the center of $AdS_3$. 
Since the spectral flow changes the energy-momentum tensor,
we need to impose the physical state condition $T^{AdS}_{\pm\pm}
+ T^{other}_{\pm\pm} = 0$ with
respect to the new energy-momentum tensor \changestr. 

\subsection{Short strings as spectral flow of timelike geodesics}

\begin{figure}[htb]
\begin{center}
\epsfxsize=2.5in\leavevmode\epsfbox{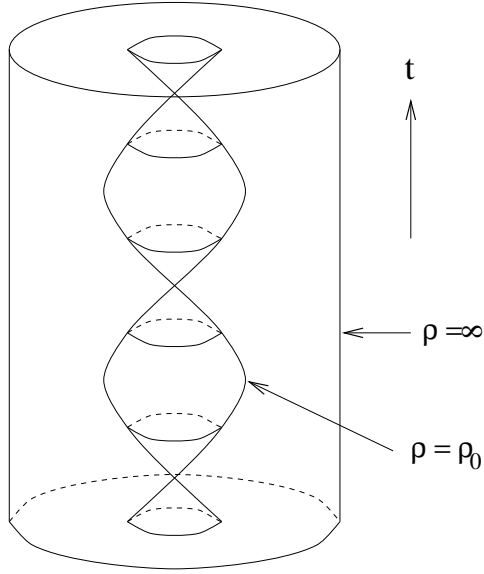}
\end{center}
\caption{A classical solution obtained by the spectral
flow of a timelike geodesic. 
The solution repeats expansion and contraction. 
The maximum size of the string is $\rho = \rho_0$.}
\label{CF1}
\end{figure}

A timelike geodesic in $AdS_3$ makes a periodic trajectory
as shown in Figure \ref{GF1}, approaching the boundary
of $AdS_3$, then coming back to the center and so on. 
In particular, when $V = U^{-1}$ in \timelikegeodesic ,
the geodesic periodically passes through the center $\rho = 0$ 
of $AdS_3$, with the period $2\pi$ in the $t$-coordinate. 
The spectral flow,
$$t \rightarrow t + w\tau,~~ \phi \rightarrow \phi + w \sigma $$ 
stretches the geodesic in the time direction and rotate
it around the center $\rho=0$; it is pictorially clear that
the resulting solution describes a circular string which
repeats expansion and contraction. This is shown 
shown in Figure \ref{CF1} in the case of $w=1$. 
Assuming $T^{other}_{\pm\pm} = h$
as in the case of geodesics, the Virasoro constraint for
the solution is
\eqn{virconstafterspectralflow}
{T_{++}^{total} = \tilde{T}_{++}^{AdS} - w \tilde{J}^3 - {k \over 4} w^2
+ T_{++}^{other} = 0. }
Since 
$$\tilde{T}^{AdS}_{++} = - {k \over 4} \alpha^2 $$ for
the timelike geodesic, we find
\eqn{energyofshortstring}
{ J_0^3 = \tilde{J}_0^3 + {k \over 2} w =
 {k \over 4} w + {1 \over w} \left( - {k \over 4} \alpha^2 + h \right).}
The spacetime energy $E$ of the string is given
by $E = 2J_0^3$, and is bounded  above as
\eqn{higherboundforshortstring}
{E= {k \over 2} w + {1 \over w} \left( - k \alpha^2 + 2h \right) 
< {k\over 2} w  + {2h \over w}.} 

It is not difficult to find an explicit form of the solution. 
When $V = U^{-1}$ in \timelikegeodesic , without loss of generality,
we can set\footnote{A different choice of
$U = V^{-1}$ simply results in shift of $\phi$ in the solution.} 
$U=V^{-1}= e^{{1\over 2}\rho_0\sigma_3}$. The 
solution\footnote{We have been informed that a similar
classical solution has also been studied 
in \cite{sanchez1,sanchez2}.}
 obtained by the spectral flow of 
(\ref{timelikegeodesic}) is then
\eqn{shortstringsolution}
 {\eqalign{ e^{i\phi} \sinh\rho 
& = i e^{iw\sigma} \sinh \rho_0~ \sin \alpha \tau  \cr
  \tan t & = {\tan w\tau + \tan \alpha \tau/\cosh\rho_0
   \over 1 - \tan w\tau  \tan \alpha \tau/\cosh \rho_0} .}}
The currents of this solution are
\eqn{currentsforshortstring}
{\eqalign{ J_R^3 & = {k \over 2} (\alpha \cosh \rho_0 + w) \cr
  J_R^\pm & = \pm i {k \over 2} \alpha \sinh \rho_0 e^{\mp i w x^+},}}
and similarly for $J_L$.
Comparing this with (\ref{energyofshortstring}), we find 
\eqn{alphaforshortstring}
{ \alpha = \alpha_\pm = - w \cosh \rho_0 \pm
\sqrt{ w^2 \sinh^2\rho_0 +{4h \over k} }.}
If we choose the branch $\alpha = \alpha_+$,
the spacetime energy $E$ of the solution is
positive and is given by 
\eqn{energyforshortstring2}
{ E = 2 J_0^3 = 2 \bar{J}_0^3 = k \left(
\cosh \rho_0  \sqrt{{4 h \over k} + w^2 \sinh^2 \rho_0} - 
w \sinh^2 \rho_0 \right) .}

There are several interesting features of this formula
for the energy $E$. Except for the case of $h = kw^2/4$, the
energy is a monotonically increasing function of
$\rho_0$, which approaches 
$E \rightarrow kw/2 + 2h/w$ as $\rho_0 \rightarrow \infty$. 
One may view that the solution describe a bound state
trapped inside of $AdS_3$. 
At the exceptional value of $h=k w^2/4$, we have $\alpha_+=0$ 
and the energy of the solution becomes $E = k w$,
completely independent of the size $\rho_0$ of the string.
The solution in this case is 
\eqn{criticalsolution}
{ \rho=\rho_0,  ~~t = w\tau,~~\phi = w \sigma, }
and represents a string staying at the fixed radius $\rho= \rho_0$,
neither contracting nor expanding. 
The fact that we have such a solution at any radius $\rho_0$
means that the string becomes marginally unstable in $AdS_3$. 

Now let us turn to the case when $U \neq V^{-1}$, or to be more 
precise,
when $U V$ does not commute with $T^3 = -{i \over 2} \sigma^2$.
(When $UV$ commutes with $T^3$, one can shift the value of 
$\tau$ to set $U = V^{-1}$.) In this case,  
the geodesic does not necessarily pass through the center
of $AdS_3$. Therefore the circular string obtained by its spectral
flow does not collapse to a point.  Since
\eqn{notequal}{\tilde{J^a}_L {T^*}^a= {k \over 2} \alpha U T^3 U^{-1},~
\tilde{J^a}_R T^a = {k \over 2} \alpha V^{-1} T^3 V,}
$\tilde J_L^3 \neq \tilde J_R^3$ 
unless $UV$ commutes with $T^3=-{i \over 2} \sigma^2$,
and the spacetime angular momentum
$ \ell = J_R^3 - J_L^3 = \tilde{J}_R^3 - \tilde{J}_L^3$
is nonzero. Thus one may view that the
circular string is kept from completely
collapsing by the centrifugal force. 
Since $T^{AdS}_{++} - T^{AdS}_{--} =  -w ( \tilde{J}_R^3
- \tilde{J}_L^3),$ the Virasoro constraint
$T_{\pm\pm}^{total}=0$ requires that the left and right
conformal weights $(h_L, h_R)$ of the internal part should
be different and that $h_R - h_L = w \ell  $.  

\begin{figure}[htb]
\begin{center}
\epsfxsize=1.7in\leavevmode\epsfbox{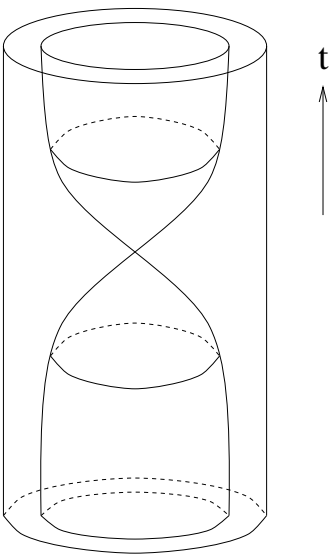}
\end{center}
\caption{A long string solution obtained by the spectral
flow of a spacelike geodesic. 
 The long string comes from the boundary 
of $AdS_3$, collapse to a point and then expands away to
the boundary of $AdS_3$ again.}
\label{CF2}
\end{figure}

\subsection{Long strings as spectral flow of spacelike geodesics}

We have seen in \higherboundforshortstring\ 
that the spacetime energy $E$ of 
the solution given by the spectral flow of
the timelike geodesic is bounded above
as $E < kw/2 + 2h / w$. What will happen
if we raise the energy above this value?
To understand this, let us look at the spectral
flow of the spacelike geodesic. Since $\tilde{T}_{++}^{AdS} = 
+ k \alpha^2/2$ for the spacelike
geodesic, the Virasoro constraint (\ref{virconstafterspectralflow})
gives
\eqn{energyoflongstring}
{ J_0^3 = \tilde{J}_0^3 + {k \over 2} w =
 {k \over 4} w + {1 \over w} \left( {k \over 2} \alpha^2 + h \right),}
and the spacetime energy is now bounded  below,
\eqn{lowerboundforlongstringenergy}
{E = 2 J_0^3 > {k \over 2} w + {2h \over w}.}

As an example, let us consider the straight line
cutting the spacelike section $t=0$ diagonally 
\simplegeodesictwo . The spectral flow with $w$
of this solution is
\eqn{simplelongstring}{
 t = w \tau, ~~\rho e^{i\phi} = \alpha \tau e^{iw\sigma},}
namely
\eqn{moresimple}{ \rho = {\alpha \over w } |t|.}
The solution starts in the infinite past $t=-\infty$ 
as a circular string of an infinite radius located at
the boundary of $AdS_3$.
The string then collapse, shrinks to a point at $t=0$, and expand 
away toward the
boundary of $AdS_3$ as $t \rightarrow + \infty$. 
More generally, if we choose $U = V^{-1} 
= e^{-{1 \over 2}\rho_0 \sigma_1}$,
the spectral flow of the geodesic \spacelikegeodesic\  
gives
\eqn{longstringsolution}
 {\eqalign{ e^{i\phi} \sinh\rho  & = e^{iw\sigma}
\cosh \rho_0 \sinh \alpha \tau  \cr
 \tan t & = { \tan w \tau + \tanh \alpha \tau \sinh \rho_0 \over
1 - \tan w \tau \tanh \alpha \tau \sinh \rho_0 }  .}}
This solution, which we call a long string,
is depicted in Figure \ref{CF2}. 

The Virasoro constraint $T_{++}^{total} = 0$ for the long string
\longstringsolution\ is 
\eqn{virasoroconstforlongstring}
{T_{++}^{AdS} + T_{++}^{other} 
= {k \over 4} \left(
  \alpha^2 - 2 \alpha w \sinh\rho_0 - w^2 \right)
+ h = 0, }
with the solutions
\eqn{alphaforlongstring}
{\alpha = \alpha_\pm =
w \sinh \rho_0 \mp \sqrt{w^2 \cosh^2 \rho_0 - {4h \over k}}.}
The spacetime energy $E$ of these solutions are
\eqn{energyoflongstring2}
{ E = 2J_0^2 = 2\bar{J}_0^3 =
 k \left( w \cosh^2 \rho_0 \mp
    \sinh \rho_0 \sqrt{w^2 \cosh^2 \rho_0 
- {4h \over k}} \right). }
At the critical value $h = kw^2/4$, 
we have $\alpha_+=0$ and
the energy for this solution becomes $E = kw$. At this point, the 
long string solution (\ref{longstringsolution}) coincides with
(\ref{criticalsolution}). Thus we see that, as we increase
the value of $h$ to $h=kw^2/4$, the short string solution
(\ref{shortstringsolution}) can turn into the long string
solution (\ref{longstringsolution}) and
escape to  infinity.

As explained in \cite{Maldacena:1998uz,Seiberg:1999xz}, a string that winds 
in $AdS_3$ close to the boundary
has finite energy because there is a balance between two large forces. 
One is the string tension that wants to make the string contract and the
other is the NS-NS $B$ field which wants 
to make the string expand. These forces
cancel almost precisely near the boundary and only a finite energy piece is
left. The threshold energy for the long string computed in 
\cite{Maldacena:1998uz,Seiberg:1999xz} is
$kw/4$, in agreement with (\ref{lowerboundforlongstringenergy})
when $h=0$.  
These strings can have some momentum in the radial direction and
that is a degree of freedom $\alpha$ that we saw explicitly above. 
One may view the long string as a scattering state,
while the previous solution (\ref{shortstringsolution}) is like a
bound state trapped inside of $AdS_3$.

In general, if $U V $ commutes with $T^3 = -{i\over 2}\sigma^2$,
the long string collapses to a point once in its lifetime. 
If $UV$ does not commute with $T^3$, the angular momentum
$\ell = J_R^3 -J_L^3$ of the solution does not vanish and the
centrifugal force keeps the string from collapsing completely. 
In this case, the Virasoro constraint $T^{total}_{\pm\pm}=0$
requires $h_R - h_L = w \ell$ for the conformal weights of
the internal sector.  

For the long strings, one can define a notion of the S-matrix. 
In the infinite past, the size of the long string is infinite
but its energy is finite. Therefore the interactions between them 
are expected to be negligible, and one can define asymptotic
states consisting of long strings. The strings then approach 
the center of $AdS_3$ and are scattered back to the boundary.   
In this process, the winding number could in principle change.

\section{Semi-classical analysis}

In studying the classical solutions, we were naively identifying the 
winding number $w$  as associated to the cycle 
$\phi \rightarrow \phi+ 2\pi$. But since this cycle is contractible
in $AdS_3$, we should be careful about what we mean by the
integer $w$.
The winding number is well-defined when the string is close to the
boundary, so we expect that long strings close to the 
boundary have definite winding numbers. On the other hand, 
when the string collapses to a point,
as shown in Figures \ref{CF1} and \ref{CF2}, the winding
number is not well-defined. Therefore, if we quantize
the string, it is possible to have a process in which
the winding number changes.
There is however a sense in which string states are characterized by
some integer $w$. 

In order to clarify the meaning of $w$ 
when the string can collapse, let us look at the Nambu action 
\eqn{nambu}{ 
S = \int dt { d\sigma \over 2 \pi} \left[ \sqrt{det g_{ind}}  - B_{t\phi}
\partial_\sigma \phi \right]
}
where $g_{ind}$ is the induced metric on the worldsheet,
and $B_{t\phi}$ is the NS-NS $B$-field. We have chosen the static 
gauge in the time direction $t=\tau$. 
We assume that initially we have a state with $\rho=0$, and we
want to analyze small perturbations. Since the coordinate
$\phi$ is not well-defined, it is more convenient to use
\eqn{cartesian}
  {X^1 + i X^2 = \rho e^{i\phi}.}
Let us compute the components of the induced metric $g_{ind}$.
To be specific, we consider the case when the target space
$AdS_3 \times S^3 \times T^4$, and consider a string winding
around a cycle on $T^4$.  
By expanding in the quadratic order in $\rho$, we find
\eqn{sec}{\eqalign{
g_{ind,00} &= k \left[ - (1 + \rho^2) + \partial_0 X^a \partial_0 X^a 
 \right] + \partial_0 Y^i \partial_0 Y^i 
\cr
g_{ind,01} & = k \partial_0 X^a  \partial_1 X^a + \partial_0 Y^i 
\partial_1 Y^i 
\cr
g_{ind,11} & = k  \partial_1 X^a \partial_1 X^a  
+ \partial_1 Y^i \partial_1 Y^i,~~~(a=1,2) }}
where $Y^i$'s are coordinates on $T^4$. For simplicity, we
consider purely winding modes on $T^4$, so that 
only $\partial_1 Y^i$ is nonzero. For these states, the conformal
weight $h$ is given by\footnote{One factor of 2 comes from the fact 
that this includes left and right
movers and the other from the fact that the expression for the energy
involves $1/2 {Y'}^2$.}
\eqn{enewind}{
 4 h = \oint {d \sigma \over 2 \pi } G_{ij} 
\partial_1 Y^i \partial_1 Y^j
}
Substituting \sec\ and \enewind\ into the action and
expanding to the quadratic order in $\rho$, we find 
\eqn{actexp}{\eqalign{
S &= \sqrt{4kh} \int d\sigma^2  \left[
1 - {1 \over 2} (\partial_0 X^a)^2 +
 {1 \over 2} {k \over 4h}
 \left( \partial_1 X^a + \epsilon_{ab} \sqrt{4h\over k} X^b\right)^2
+ \cdots \right]
 \cr 
& = \sqrt{4kh} \int d\sigma^2 
\left[
1 - {1 \over 2} |\partial_0 \Phi|^2 +
 {1 \over 2} {k \over 4h}
 \left| \left(\partial_1 -i \sqrt{4h\over k}\right) \Phi \right|^2
+ \cdots \right] ,
}}
where $\Phi = X^1 + i X^2$. 

The action (\ref{actexp}) is the one for a massless charged scalar
field on $\IR \times S^1$ coupled to a constant gauge field 
$A = \sqrt{4h/k}$ around $S^1$. As we vary $A$, we observe 
the well-know phenomenon of the spectral asymmetry. 
Let us first  assume that $A$ is not an integer. 
A general solution to the equation of motion derived from
(\ref{actexp}), requiring the periodicity in $\sigma$, is 
\eqn{generalquad}
{ \Phi \sim  \sum_{n=-\infty}^{\infty}
\left( a_n^\dagger  e^{i(n-A)(\tilde{\tau} + \sigma)}
          + b_n e^{-i(n-A)(\tilde{\tau} - \sigma)} \right)
     {e^{iA \sigma}  \over n - A} ,}
where $A = \sqrt{4h/k}$ and $\tilde{\tau} = \tau/A$.
Upon quantization, the commutation relations are given (modulo
a positive constant factor) by
\eqn{abcommutator}
{ [ a_n, a_m^\dagger ] = (n-A) \delta_{n,m},~~~
  [b_n, b_m^\dagger ] = (n-A) \delta_{n,m}.}
Notice that the sign in the right hand side of \abcommutator\ 
determines whether $a_n$ or $a_n^\dagger$ should be regarded as
the annihilation operator. 
Thus, assuming that the Hilbert space is positive definite,
the vacuum state is defined by
\eqn{vacuumasym}
{\eqalign{ & a_n | 0 \rangle = b_n |0 \rangle = 0,  ~~~(n > A) \cr
 & a_n^\dagger | 0 \rangle = b_n^\dagger | 0 \rangle = 0,
  ~~~(n < A).}}
For $\Phi = \rho e^{i\phi}$ given by (\ref{generalquad}) and 
$t = A \tilde{\tau}$, we find
\eqn{currentnearzero}
{\eqalign{ J_R^+ & =  k \left( e^{-i t} \partial_+ \Phi^*
          - \Phi^* \partial_+ e^{-i t} \right) 
\sim  -ik \sum_n   a_n e^{-in  (\tilde{\tau} + \sigma)} \cr
J_R^- & =  k \left( e^{i t} \partial_+ \Phi
          - \Phi \partial_+ e^{i t} \right) 
\sim ik \sum_n a_n^\dagger e^{in  (\tilde{\tau} + \sigma)} ,}}
and similarly for $J_L^\pm$.
Therefore $J_n^+ = - ik a_n$ and $J_n^- = ik a_{-n}^\dagger$. 
The vacuum state $|0 \rangle$ defined by (\ref{vacuumasym})
then obeys
\eqn{gaussianspectralflow}
{ J_n^+ |0 \rangle = 0~~~(n>A),~~~J_n^- | 0 \rangle =0 ~~~ (n > -A). }
Thus the vacuum state $|0\rangle$ is not in a regular
highest weight representation of the current algebra
$\widehat{SL}(2,R)$. If we set 
\eqn{changecur3}
{J_n^\pm = \tilde{J}_{n\mp w}^\pm}
with the integer $w$ defined by
\eqn{jumpformula}
{   w < A < w+1,}
then $|0 \rangle$ obeys the regular highest weight condition
with respect to $\tilde{J}_n^\pm$,
\eqn{usualcondition}{
\tilde{J}_{n}^+ | 0 \rangle 
= 0 ~~(n \geq 1), ~~~~~\tilde{J}_{n}^- | 0 \rangle = 0 
~~ (n \geq 0) } 
The change of the basis (\ref{changecur3})
is nothing but the spectral flow (\ref{changecur2}) discussed earlier,
so we can identify $w$ as the amount of spectral flow needed to 
transform the string state into a string state which obeys the
regular conditions \usualcondition . 
We have found that, for a given value of $h$,
there is a unique integer of $w$ associated to the string state.
As we vary the conformal weight $h$,  $A = \sqrt{4h/k}$
will become an integer. At that point, one of the modes of the 
field $\Phi$ will have a vanishing potential. In fact we can
check that classically this potential is completely flat. Giving an
expectation value to that mode, we find configurations
as in  \criticalsolution. Corresponding to various 
values of its momentum in the radial direction, 
we have a continuum of states.
So, at this value of $h$, we do not have a 
normalizable ground state; instead we 
have a continuum of states which  are $\delta$-function 
normalizable.
If we continue to increase $h$, we find again 
normalizable states, but they are labeled by a new integer 
$(w+1)$.  Notice that $w$ is not directly related to the physical
winding of the string. In fact by exciting a coherent state of 
the oscillators $a_n$ or $b_n$ we can find string states that
look like expanding and collapsing strings with winding number $n$ around
the origin.

One of the puzzles we raised in the introduction was what
happens when we increase the internal conformal weight $h$ 
of the string beyond the upper bound implied by the restriction 
$ j < k/2$ on the $SL(2,R)$ spin $j$ due to the no-ghost theorem. 
In this section, we saw
a semi-classical version of the puzzle and its resolution.
When $h$  reaches the bound, we find that the state can become 
a long string with no cost in energy. Above the bound, 
we should consider a Fock space  with
a different bose sea level. 
In the fully quantum description of the model given below, 
we will find a similar situation but with 
minor corrections.

\section{Quantum string in AdS$_{\bf 3}$}

The Hilbert space of the WZW model is a sum of products 
of representations of the left and the right-moving current
algebras generated by
\eqn{generators}
{  J_L^a = \sum_{n=-\infty}^\infty J_n^a e^{-inx^-},
  ~~ J^a_R =\sum_{n=-\infty}^\infty \bar{J}_n^a e^{-inx^+},}
with $a=3, \pm$, obeying the commutation relations
\eqn{comm}{
\eqalign{ &[J^3_n, J^3_m ]  = - {k \over 2} n \delta_{n+m,0} \cr
          &[J^3_n, J^\pm_m ]  = \pm J^\pm_{n+m} \cr
          &[J^+_n , J^-_m ] = -2J^3_{n+m} + kn\delta_{n+m,0} ,}}
and the same for $\bar{J}_n^a$. We denote the current
algebra by $\widehat{SL}_k(2,R)$. 
The Virasoro generator $L_n$ are defined by
\eqn{sugaw}{
\eqalign{ &L_0  = {1 \over k-2}\left[ {1 \over 2} (J_0^+ J_0^-
+ J_0^- J_0^+) - (J_0^3)^2 + \sum_{m=1}^\infty
(J_{-m}^+ J_{m}^- + J_{-m}^- J_{m}^+  - 2 J_{-m}^3 J_{m}^3)
\right] \cr &
L_{n\neq 0} =
{1 \over k-2} \sum_{m=1}^\infty
(J_{n-m}^+ J_{m}^- + J_{n-m}^- J_{m}^+  - 
  2 J_{n-m}^3 J_{m}^3) }}
and obey the commutation relation
\eqn{vircom}{ [L_n, L_{m}] =
          (n-m) L_{n+m} + {c \over 12} (n^3-n) \delta_{n+m,0},}
where the central charge $c$ is given by
\eqn{central}{ c = {3k \over k-2}. }

We will find that the Hilbert space of the WZW model
consists of sub-sectors parameterized by integer $w$, labeling
the amount of spectral flow in a sense to be made precise below.
We then formulate our proposal on how the complete Hilbert 
space of the WZW model 
is decomposed into representations of the current
algebras and provide evidences
for the proposal. 

States in a representation of the current algebra are labeled by
eigenvalues of $L_0$ and $J_0^3$. Since the kinetic term of
the WZW model based on $SL(2,R)$ has an indefinite signature, 
it is possible that the Hilbert space of the model contains
states with negative eigenvalues of $L_0$ as well as states
with negative norms, and indeed both types of states appear
as we will see below. For the moment, we will  consider 
a representation in which eigenvalues of $L_0$ is 
bounded  below. We call them {\it positive energy 
representations}, or {\it unflowed} representations.
 Since the action of $J_n^{3,\pm}$ with $n \geq 1$ 
on a state lowers the eigenvalue of $L_0$ by $n$,
 there has to be a set of states 
which are annihilated by them. We will call such states the  
primary states of the positive energy representation. 
All other states in the
representation are obtained by acting $J_{-n}^{3,\pm}$
($n \geq 1$) on the primary states. The ground
states make a representation of $SL(2,R)$
generated by  $J_0^{3,\pm}$. So
let us review irreducible representations of $SL(2,R)$. 

\subsection{Representations of the zero modes}

We expect that physical states of a string in $AdS_3$
 have positive norms. Since $J_0^{3,\pm}$ commute with 
the Virasoro constraints, physical spectrum of the string
must be in unitary representations of $SL(2,R)$.
Most of the mathematical references on representation theory 
of $SL(2,R)$ deal with the case with compact time\footnote{For 
a review of  representations of $SL(2,R)$,
see for example \cite{sugiura}.};
 we are however interested in 
the case with non-compact time. A clear analysis from the algebraic
point of view is presented in \cite{Dixon:1989cg}, which we now 
summarize with some minor changes is notation.

There are the following five types of unitary 
representations.
All the representations are parameterized by $j$,
which is related to the second Casimir
$c_2 =  {1 \over 2}(J_0^+ J_0^- + J_0^- J_0^+) - (J_0^3)^2$
as $c_2 = - j(j-1)$. 

\medskip
 
\noindent 
(1) Principal discrete representations (lowest weight):

A representation of this type is realized in
the Hilbert space $${\cal D}_j^+= \{
|j; m\rangle :~ m=j, ~j+1, ~j+2, \cdots \},$$
where $|j; j \rangle$ is annihilated by $J_0^-$
and $|j; m \rangle$ is an eigenstate of $J_0^3$
with $J_0^3 = m$.
The representation is unitarity if $j$ is real and $j > 0$.
For representations of the group 
$SL(2,R)$, $j$ is restricted to be a half of integer. 
Since we are considering the universal cover of $SL(2,R)$,
$j$ can be any positive real number. 

\smallskip

\noindent
(2) Principal discrete representations (highest weight):

A charge conjugation of (1). 
A representation of this type is realized in
the Hilbert space $${\cal D}_j^-= \{
|j; m\rangle :~ m= -j, ~-j-1, ~-j-2, \cdots \},$$
where $|j; j\rangle$ is annihilated by $J_0^+$
and $|j; m \rangle$ is an eigenstate of $J_0^3$
with $J_0^3 = m$. 
The representation is unitary if $j$ is real and $j > 0$. 

\smallskip

\noindent
(3) Principal continuous representations:

A representation of this type is realized
in the Hilbert space of $${\cal C}_j^{\alpha}
= \{  |j ,\alpha; m \rangle :~m = \alpha,~ \alpha \pm 1, 
\alpha \pm 2,~ \cdots \},$$
where $|j,\alpha; m \rangle$ is an eigenstate
of $J_0^3$ with $J_0^3=m$. 
Without loss of generality,
we can restrict $0 \leq \alpha < 1$. 
The representation is unitary if $j = 1/2 + i s$
and $s$ is real\footnote{Strictly speaking the representation with
$j=1/2, ~\alpha=1/2$ is  reducible as the sum of a highest weight and
a lowest weight representation with $j=1/2$.}. 

\smallskip

\noindent
(4) Complementary representations:

A representation of this type is realized
in the Hilbert space of $${\cal E}_j^{\alpha}
= \{  |j ,\alpha; m \rangle :~  m = \alpha,~ \alpha \pm 1, 
~\alpha \pm 2, \cdots \},$$
where $|j,\alpha; m \rangle$ is an eigenstate
of $J_0^3$ with $J_0^3=m$. 
 Without loss of generality,
we can restrict $0 \leq \alpha <1$. 
The representation is unitary if $j$ is  real, with 
$1/2 < j < 1 $ and  $ j-1/2 < |\alpha -1/2|$.

\smallskip
\noindent
(5) Identity representation:

This is the trivial representation with $j=0$. 

\medskip

The analysis that led to the above representation was completely algebraic
and in a particular physical system we can have only a subset of all 
possible representations. 
Which of these representations appear in the
Hilbert space of the WZW model? As the first approximation,
let us consider the $k \rightarrow \infty$ limit.
If we expand around a short string solutions, $i.e.$
oscillations near geodesics in $AdS_3$, 
the WZW model in this limit reduces to the quantum 
mechanics on $AdS_3$. The Hilbert space
of the quantum mechanical model is the space of 
square-integrable\footnote{Since $AdS_3$ 
is non-compact, we consider 
square-integrability in the delta-function sense.}
 functions ${\cal L}^2(AdS_3)$ on $AdS_3$.   
The isometry of $AdS_3$ is 
$SL(2,R) \times SL(2,R)$, and one can decompose
${\cal L}^2(AdS_3)$ into its unitary representations.
It is convenient to choose the basis of the Hilbert space
in the following way.  For each representation
${\cal R}$, one can define a function on $AdS_3$ by 
$F_{m, \bar{m}}(g) = \langle m | g | \bar{m} \rangle$
where $g \in AdS_3$, $i.e.$ universal cover of $SL(2,R)$, 
and $|m \rangle$ is an eigenstate of $J_0^3$ with $J_0^3=m$. 
Thus, for a given representation ${\cal H}$ of
$SL(2,R)$, the function $F_{m,\bar{m}}(g)$
on $AdS_3$ is in the tensor product of the representations
${\cal R} \times {\cal R}$
for the isometry group $SL(2,R) \times SL(2,R)$.
%

For a discrete representation ${\cal D}_j^\pm$,
the wave-function $f(\rho)$ behaves as $f(\rho)
\sim e^{- 2 j \rho}$ for large $\rho$. Thus 
$\phi \in {\cal L}^2(AdS_3)$ if  $j > 1/2$.
Notice that in the range $0<j<1$ we have two representations with
the same value of the Casimir but only one is in ${\cal L}^2(AdS_3)$,
the one with $ 1/2 < j<1$.
As explained in  \cite{Klebanov:1999tb}, one could modify the norm
so that the second solution with $0<j<1/2$ becomes normalizable. 
This modification of the norm is $j$-dependent. Similarly, supplementary 
series representations need a $j$-dependent modification to the 
norm to render them normalizable \cite{sugiura}. Therefore these 
representations would appear in non-standard quantizations 
of geodesics,
quantizations which do not use the ${\cal L}^2$ norm on $AdS_3$. 
In this paper, we will only consider the standard quantization
using the ${\cal L}^2$ norm for the zero modes\footnote{
Notice however, that even if the primary states 
 have $j>1/2$, we could have states with smaller values of $j_0$ for the 
zero mode $SL(2,R)$ among the descendents,  for example 
$J^-_{-1} | j\rangle$ with 
$ 1<j<3/2$, has $j_{0} = j-1 <1/2$.}.
Wave-functions in ${\cal C}^\alpha_{j=1/2+is}$ are 
also delta-function normalizable with respect to the ${\cal L}^2$ norm. 
It is known that ${\cal C}^\alpha_{j=1/2+is} \times 
{\cal C}^\alpha_{j=1/2+is}$ and ${\cal D}_j^\pm \times {\cal D}_j^\pm$ 
with $j > 1/2$ form the complete basis of ${\cal L}^2(AdS_3)$. 

For discrete lowest weight representations, 
the second Casimir is bounded  above 
as $c_2=-j(j-1) \leq 1/4$. 
This corresponds to the well-known
Breitenlohner-Freedman bound (mass)$^2 \geq -1/4$
for the Klein-Gordon equation.
For the principal continuous representation
${\cal C}_j^\alpha$ with $j=1/2 + i s$,
the second Casimir is $c_2 =  1/4 + s^2$.
Therefore an existence of such a particle would
violate the Breitenlohner-Freedman bound. 
In the bosonic string theory, the only physical
state of this type is the tachyon. In 
a perturbatively stable string theory, such
particle states should be excluded from its
physical spectrum.
On the other hand, 
the continuous representations 
appear in ${\cal L}^2(AdS_3)$
and they
are expected to be part of the Hilbert space of the WZW 
model before the Virasoro constraint is imposed.

\subsection{Representations of the current algebra and no-ghost theorem}

\begin{figure}[htb]
\begin{center}
\epsfxsize=3.5in\leavevmode\epsfbox{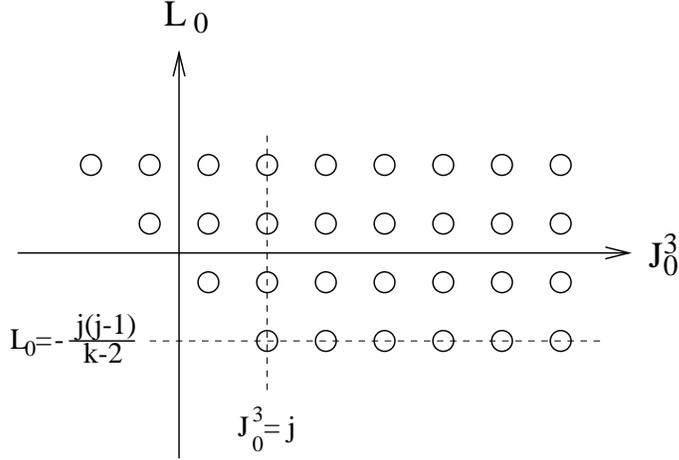}
\end{center}
\caption{Weight diagram the representation  $\widehat {\cal D}^{+}_j$, 
whose the primary states form a  discrete lowest weight representation
${\cal D}_j^+$.}
\label{F1}
\end{figure}

Given a unitary representation ${\cal H}$ of
$SL(2,R)$, one can construct a representation of $\widehat{SL}(2,R)$
by regarding ${\cal H}$ as its primary states annihilated
by $J_{n\geq 1}^{3,\pm}$.  The full representation space 
is generated by acting $J_{n \leq -1}^{3,\pm}$ on ${\cal H}$.
Following the discussion in the previous subsection,
we consider the cases when ${\cal H} = {\cal C}_{j=1/2 + is}^\alpha$
and ${\cal D}_j^\pm$ with $j > 1/2$. 
We denote by $\widehat {\cal D}^{\pm}_j$ and $\widehat {\cal C}^{\alpha}_j$
the representations of the full current algebra built on the 
corresponding representations  of the zero modes. 
In Figure \ref{F1},
we have shown the weight diagram of the positive energy
representation $\widehat {\cal D}^{+}_j$.

A representation of $\widehat{SL}_k(2,R)$ in general 
contains states with negative norms. In order for a string 
theory on $AdS_3$ to be consistent,
one should be able to remove these negative norm 
states by imposing the Virasoro constraint,
\eqn{virconst}{ (L_n + {\cal L}_n - \delta_{n,0}) 
   |{\rm physical}\rangle = 0, ~~n \geq 0,}
on the Hilbert space for a single string state,
where $L_n$ is the Virasoro generator of the
$SL(2,R)$ WZW model and ${\cal L}_n$ for the
sigma-model on ${\cal M}$. It has been shown that
this no-ghost theorem holds  for 
states  in $\widehat {\cal C}_{j=1/2 + i s}^{\alpha}$
or $\widehat {\cal D}_{j}^\pm$ with $0 < j < k/2$ 
\cite{Balog:1989jb,Petropoulos:1990fc,%
Hwang:1991aq,Hwang:1992an,Hwang:1998tr,Evans:1998qu,Dixon:1989cg}. 

The no-ghost theorem is proved by first showing
that all the solutions to the Virasoro constraint 
(\ref{virconst}) can be expressed, modulo null states,
as states in the coset $SL(2,R)/U(1)$ obeying
\eqn{coset}{ J_n^3 |\psi \rangle = 0~,~~~~~~~~~n\geq 1. }
This statement is true for $\widehat{\cal C}_{1/2+is}^\alpha$ 
and $\widehat{\cal D}_j^\pm$ with $0 < j < k/2$, if the total central
charge of the Virasoro generator $L_n + {\cal L}_n$
is $26$
\cite{Balog:1989jb,Petropoulos:1990fc,Mohammedi:1990dp,%
Hwang:1991aq,Hwang:1992an,Hwang:1998tr,Evans:1998qu}
\footnote{We also assume $k > 2$.}.
We review the proof of this statement in the appendix A.1.
The second step is to show that the condition
(\ref{coset}) removes all negative norm states.
This was shown in \cite{Dixon:1989cg} for 
the same class of representations. 

The no-ghost theorem suggests that the spectrum of
discrete representations has to be truncated for
$j < k/2$. As we will see, this truncation is closely
related to the existence of the long string states. 

\subsection{Spectral flow and the long string}

The classical and semi-classical results discussed above 
indicate that, beyond
 positive energy representations that we have
discussed so far, we have to include others related by 
spectral flow.
To define a quantum version of the
spectral flow, we note that, for any integer $w$,
the transformation
$J_n^{3,\pm} \rightarrow \tilde{J}_n^{3,\pm}$ given by 
\eqn{flow}{
  \tilde{J}_n^3 = J_n^3 - {k \over 2} w \delta_{n,0},
     ~~ \tilde{J}_n^+ = J_{n+w}^+, ~~ \tilde{J}_n^-
 = J_{n-w}^-,}
preserves the commutation relations (\ref{comm}).
The Virasoro generators $\tilde{L}_n$, which 
have the standard Sugawara form in terms of   $\tilde{J}_n^{a}$,
are different from $L_n$. They are given by 
\eqn{virflow}{  \tilde{L}_n = L_n + w J_n^3 - {k \over 4} w^2 \delta_{n,0}.}
Of course, they obey the Virasoro algebra with the same central charge $c$.
This is the same formula as saw in the classical counterpart
(\ref{changestr}) of the spectral flow.

\begin{figure}[htb]
\begin{center}
\epsfxsize=3.8in\leavevmode\epsfbox{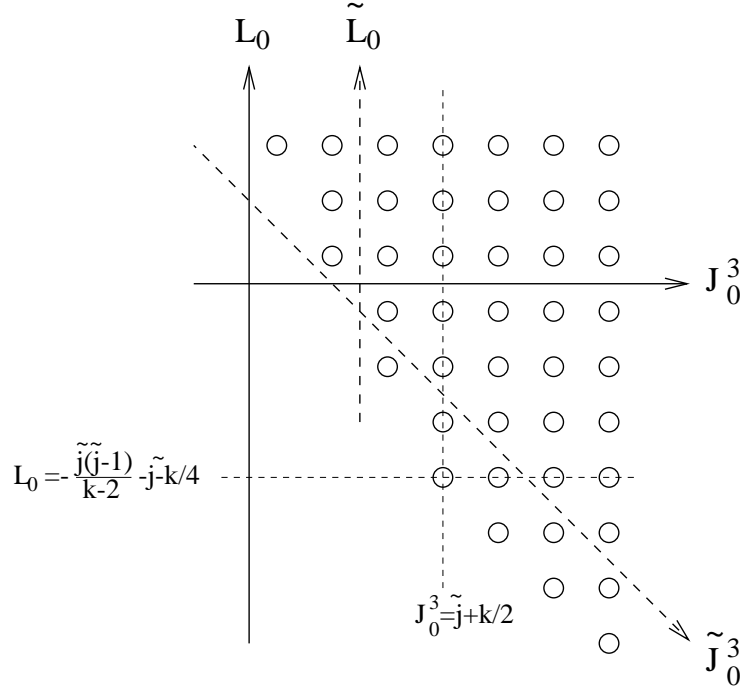}
\end{center}
\caption{Weight diagram of the representation $\widehat {\cal
D}^{+,w=1}_{\tilde j}$, which is the spectral flow of the diagram \ref{F1} with
$w=1$. The worldsheet energy $L_0$ of this representation
is not bounded below, but the spacetime energy, $J^3_0$, is bounded
below for states obeying the Virasoro constraint $L_0=1$.}
\label{F2}
\end{figure}

\begin{figure}[htb]
\begin{center}
\epsfxsize=3.8in\leavevmode\epsfbox{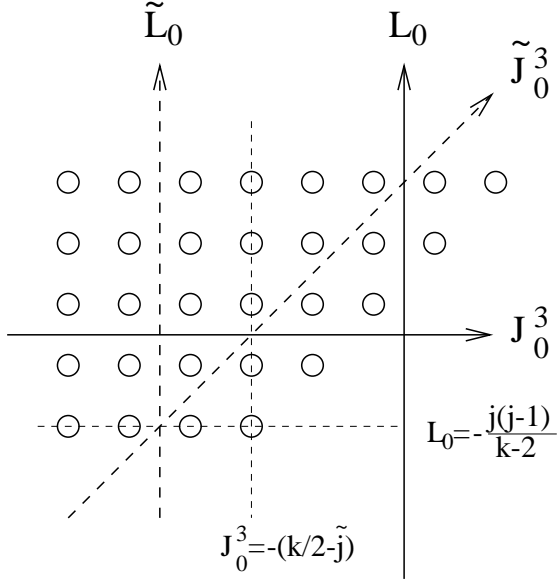}
\end{center}
\caption{The spectral flow of the diagram \ref{F1} with
$w=-1$. $\widehat{\cal D}_{\tilde{j}}^+$ is mapped to 
$\widehat{\cal D}_{\tilde j}^{+, w=-1} = \widehat {\cal D}^{-}_{ j}$
with $j = k/2-\tilde{j}$. Since $\tilde{j} > 1/2$, 
the resulting $\widehat {\cal D}_j^-$ obeys $j <  (k-1)/2$.
In particular, the unitarity bound $j < k/2$ required
by the no-ghost theorem is satisfied.}
\label{F3}
\end{figure}

The change of the basis (\ref{flow}) maps one representation
into another, and this is called the spectral flow.
In the case of a compact group such as $SU(2)$, the spectral
flow maps a positive energy representation 
of the current algebra into another positive energy 
representation. An analogous transformation in the case
of the $N=2$ superconformal algebra in two dimensions
has been used to construct the spacetime supercharges
for superstring. 

In the case of $SL(2,R)$, the spectral flow generates
a new class of representations.
As shown in Figure \ref{F2},
the spectral flow with $w=1$ maps the lowest weight
representation $\widehat{\cal D}_{\tilde{j}}^+$ to a representation
in which $L_0$ is not bounded  below. The appearance
of negative energy states is not too surprising since the kinetic
term of the $SL(2,R)$ model is not positive definite. 
In general, a spectral flow of $\widehat{\cal D}_{\tilde{j}}^+$
with $w\geq 1$ or $w \leq -2$ gives a new representation
in which $L_0$ is not bounded below.  
Similarly, the spectral flow of $\widehat{\cal C}_{j=1/2+is}^\alpha$
with $w \neq 0$ gives a representation in which $L_0$
is not bounded below. 
We denote the resulting representations by $\widehat{\cal D}^{\pm
,w}_{\tilde j}$ and $\widehat {\cal C}^{\alpha, w}_{\tilde j}$, where
$\tilde j$ labels the $SL(2,R)$ spin before the spectral flow.

These representations obtained by the spectral flow also
contain negative norm states. 
In Appendix A.2, we generalize the proof of the no-ghost theorem
and show that the Virasoro constraints indeed remove all
negative norm states in 
the representations $\widehat{\cal C}_{j=1/2 + i s}^{\alpha, w}$
and $\widehat{\cal D}_{\tilde j}^{\pm , w}$ with $\tilde j < k/2$,
for any integer $w$. 

The only case where we get a representation with $L_0$ bounded below
by the spectral flow is 
$\widehat{\cal D}_j^\pm$ with $w=\mp 1$. In this case, 
the representation is mapped to another positive energy
representation $\widehat{\cal D}^{\pm,w=\mp 1}_{\tilde j} =
\widehat{\cal D}_{k/2-\tilde{j}}^\mp$.
Note that, if we start with
the representation with $\tilde{j} > 1/2$, the representation
one gets after the spectral flow satisfies 
$j = k/2 - \tilde{j} < (k-1)/2$.
Conversely, if there were a representation $\widehat{\cal D}_j^\pm$
with $j > (k-1)/2$ in the Hilbert space, the spectral flow would
generate a representation $\widehat{\cal D}_j^\mp$ with
$j < 1/2$, in contradiction with the standard harmonic analysis
of the zero modes in section 4.1.
Therefore, if we assume that the spectral flow
is a symmetry of the WZW model,
the discrete representations $\widehat {\cal D}_j^\pm$ appearing
in the Hilbert space are automatically 
restricted to be in $1/2 < j < (k-1)/2$. In particular, 
the spectrum of $j$ is truncated below 
the unitarity bound $j < k/2$ required by the no-ghost 
theorem. This further restriction on $j$  was discussed in a related
context by \cite{Giveon:1999px}.

\subsection{Physical spectrum}

Let us consider first the spectrum for strings with $w=0$. 
This is fairly standard. We start from an arbitrary descendent
at level $N$ in the current algebra and some operator of the internal
CFT with conformal weight $h$. 
The $L_0$ constraint reads
\eqn{lzeroc}{
(L_0 -1)|j,m,N,h\rangle =0 \Longrightarrow 
 -{  j (j-1) \over k-2 } + N + h -1 =0
}
If we demand that $ 1/2 \leq j \leq (k-1)/2$,
this equation will have a solution as long as $N +h$ is within the range
\eqn{range}{
0 \leq  N + h -1 + { 1 \over 4(k-2)}\leq  {(k-2) \over
4 } 
}
If we allow $j$ to go all the way to $k/2$ we get $k/4$ on  the right
hand side of \range .

To analyze physical states of strings with $w\neq 0$,
we start with a positive energy representation 
$\widehat {\cal D}_{\tilde{j}}^+$.
After the spectral flow (\ref{flow}), a primary state
$|\tilde{j}, \tilde{m}\rangle$ of $\widehat {\cal D}_{\tilde{j}}^+$,
as a state of $\widehat {\cal D}^{+,w}_{\tilde j}$,  obeys 
\eqn{afterflow}{
\eqalign{ &   J^+_{n+w} | \tilde{j}, \tilde{m} \rangle = 0,~~
 J^-_{n-w} | \tilde{j}, \tilde{m} \rangle = 0, ~~
J^3_n | \tilde{j}, \tilde{m} \rangle = 0,
                 ~~~~ n \geq 1 \cr
          &
   J_0^3  | \tilde{j}, \tilde{m} \rangle
    =  \left({k \over 2} w + \tilde{m}\right) 
| \tilde{j}, \tilde{m} \rangle .}}
Let us look for physical states with respect to the Virasoro
generator $L_n$. From (\ref{afterflow}), we find
the Virasoro constraints are
\eqn{phys}{
\eqalign{& (L_0 -1) | \tilde{j}, \tilde{m} \rangle
= \left( -{ \tilde{j}(\tilde{j}-1) \over k-2} -  w \tilde{m}
   - {k \over 4} w^2  + \tilde{N} +h -1 \right)
| \tilde{j}, \tilde{m}, \tilde{N}, h \rangle
=0 \cr
 & L_n | \tilde{j}, \tilde{m} \rangle
= (\tilde{L}_n - w \tilde{J}_n^3 ) | \tilde{j}, \tilde{m} \rangle
 = 0,~~~~~~ n \geq 1 .}}
where $h$ is the contribution to the conformal weight from the internal 
CFT and $\tilde{N}$ is the level inside the current algebra
before we take the spectral flow. 
The state  obeys the physical state
conditions provided
\eqn{onshell}{
 \tilde{m} =  -{k \over 4} w  + { 1 \over w} \left(
- {\tilde{j}(\tilde{j}-1) \over (k-2) } +
                  \tilde{N} + h -1 \right) . }
The spacetime energy of this state measured by $J_0^3$ is then
\eqn{energyofboundstring}{
J_0^3 = \tilde m  + {k \over 2} w =
 {k \over 4} w  + { 1 \over w} \left(
- {\tilde{j}(\tilde{j}-1) \over (k-2) } +
                  \tilde{N} + h -1 \right) . }
This is the quantum version of the classical formula
(\ref{energyofshortstring}), with the replacement
$${k \over 4} \alpha^2 \rightarrow {\tilde{j}(\tilde{j}-1) \over k-2} 
+ 1.$$ 
Notice that $\tilde m = \tilde j +q$ where $q$ is some integer, which could
be negative\footnote{$\tilde m$ is the total $\tilde J^3$ eigenvalue of the 
state so it can be lowered by applying $J^-_{-n} $ to the highest weight
state. So we have the constraint $q \geq -\tilde{N}$.}.
 Therefore the physical state condition becomes
\eqn{spectfl}{
\tilde j = { 1 \over 2} - { k-2 \over 2} w + \sqrt{ { 1\over 4} + 
(k-2) \left( h -1 + N_w - { 1 \over 2} w(w+1) \right)}.
}
Here
\eqn{ntilde}{
  N_w = \tilde{N} - wq}
is the level of the current algebra after the spectral flow by
the amount $w$. 
Notice that the equation for $\tilde j$ is invariant under 
$\tilde{N} \to \tilde{N} \pm w $, 
$ q \to q \pm 1 $. This is reflecting the fact that $J^\pm_0 = \tilde
J^\pm_{\mp w} $ commute with the Virasoro constraints and generate
the spacetime $SL(2,R)$ multiplets. 
In particular,  we see that
the spacetime 
 $SL(2,R)$ representations that we get are lowest energy representations,
since repeated action of 
 $J_0^- = \tilde J^-_w$ will eventually  annihilate the state.
In fact, it is shown in Appendix A.2 that 
 the only physical state with zero spacetime energy,
$J^3_0=0$, is the state $J^-_{-1} |j =1\rangle$, and its complex conjugate.
This physical state corresponds to the dilaton field
in $AdS_3$, which played an important role in
the analysis of the spacetime Virasoro algebra
in \cite{Kutasov:1999xu}. 
All other states (except the tachyon with $ w =0$) have nonzero energy,
and form highest/lowest weight representations of $SL(2,R)$ spacetime
algebra. The negative energy ones are the complex conjugates of the
positive energy ones.

By solving the on-shell condition (\ref{spectfl})
for $\tilde{j} > 0$ and substituting it into
(\ref{energyofboundstring}), one finds that the
spacetime energy of the string is given by
\eqn{discretespectrum}{
 {E + \ell \over 2} = J_0^3 = 
 q + w + {1 \over 2} 
 + \sqrt{{1 \over 4} + (k-2)\left(h-1+ N_w - {1\over 2} w(w+1)\right)}. }
Since both $N_w$ and $q$ are integers,
the energy spectrum is discrete. 
This is reasonable since we are considering the string
trapped inside of $AdS_3$.
The constraint $1/2 < \tilde{j} <(k-1)/2$ translates into
the inequality
\eqn{rangenew}
{{k \over 4} w^2 + {w \over 2}
< N_w + h -1  + {1 \over 4(k-2)}  <  
{k \over 4} (w+1)^2 - {w+1 \over 2}.}
This is the quantum version of the semi-classical formula
(\ref{jumpformula}). In fact, if we take $k, h \gg \tilde{N}, q, w$,
(\ref{rangenew}) reduces to (\ref{jumpformula}). 
As in the semi-classical discussion, $w$ is not necessarily related to the
physical winding number of the string. It is just an integer labeling the 
type of representation that the string state is in.

The analysis for the representations coming from the continuous 
representations for the zero modes is similar. 
If we do not spectral flow, the only state in the continuous 
representation is the tachyon. 
If we do spectral flow, we get the equation \onshell , which 
can be conveniently rewritten as 
\eqn{onshcon}{
J_0^3 = \tilde m + {w k \over 2} = { k w \over 4} + 
{ 1 \over w } \left(
{ { 1 \over 4} + s^2 \over k-2} + \tilde{N} + h -1 \right)
}
For continuous representations $w$ is labeling the physical winding
of the string when it approaches the boundary of $AdS$. 
In this case we do not get an equation like \spectfl\ since,
for continuous representations,  
$\tilde m$ is not related to $j$. 
Comparing with the classical formula
(\ref{energyoflongstring}), we
identify $s$ as the momentum $\alpha/k$ of the long string
along the radial direction of $AdS_3$.
We clearly see that the energy of this state is above the threshold to 
produce an excitation that will approach the boundary as a $w$-times wound
string.

We can see that, whenever the value of $h$ is such that it
saturates the range \rangenew , we have a continuous representation 
with the same energy. This is clear for the lower bound in
the case of $w=0$ since, for each state in the discrete 
representation with $j=1/2$, there is one in the continuous 
representation with the same values of $L_0$ and
$J_0^3$. By the spectral flow, we see that the same is true for the
lower bound in \rangenew\ for any $w$. Indeed we can
check explicitly that a state in the discrete representation with
parameters ($h,w,q,\tilde{N}$) saturating the lower bound
in \rangenew\ has the same spacetime energy as a state in the 
continuous representation with parameters ($h,w,s=0,\tilde{N}$).
(The parameter $\alpha$ in the continuous representation
is fixed by the value of $J_0^3$ in (\ref{onshcon}).)
Similarly, if we have a state in a  discrete representation
saturating the upper bound in \rangenew , it has the
same spacetime energy as a state in the continuous representation
with parameters ($h,w+1,s=0, \tilde{N}'=\tilde{N}+q$). 
Note that, since $q\geq -\tilde{N}$ (see
the footnote in the previous page), we have $\tilde{N}' \geq 0$. In this case,
to show that the two states have the same energy, it is useful to 
identify the state in ${\cal D}^{+,w}_{\tilde j = \tilde j}$ 
as a state in ${\cal D}^{-,w+1}_{\hat j = k/2-\tilde j}$.
Since $\tilde j \to (k-1)/2$ corresponds to $\hat j \to 1/2$
under this identification, we can apply the above argument for 
the lower bound to show that we will find a state in the continuous 
representation. The shift $\tilde{N}' = \tilde{N} + q$ comes 
from the fact that 
the identification ${\cal D}^{+,w}_{\tilde j = \tilde j}= 
{\cal D}^{-,w+1}_{\hat j = k/2-\tilde j}$ involves 
spectral flow one more time. 

The above paragraph explains what happens as we change $\tilde{j}$ 
in a discrete representation and we make it equal to the upper or lower 
bound: a continuous representation appears. 
Another question that one could ask is the following. Given a value of $h$, 
what is the state with the lowest value of $J^3_0$ that satisfies the
physical state conditions? 
Let us first look for the lowest energy state in the
discrete representations obeying the bound \rangenew .
Within this bound, one can show that 
$\partial J_0^3(h,w,q,\tilde{N})/\partial q \geq 0$ 
and $\partial J_0^3(h, w, q=-\tilde{N}, \tilde{N})/\partial \tilde{N} \geq 0$.
Therefore, if we can set $q=\tilde{N}=0$,
it will give the lowest energy state in the discrete
representations. This is possible if $h$ is within the range, 
\eqn{rangeh}{
 {k \over 4} w^2 + { w \over 2}  < h-1 + { 1 \over 4 (k-2)} < { k \over 4}
(w+1)^2 - { w+1 \over 2}.
}
With some more work, one can show that, for $h$ in this 
range, there isn't any state in a continuous representation whose
energy is lower than that of the discrete representation state
with $\tilde{N}=q=0$. As we saw in the above paragraph, at the
upper or lower bound of \rangeh , the energy of the discrete state
$(q=0, \tilde{N}=0)$ coincides with that of the continuous state with 
$(s=0, \tilde{N}=0)$. Outside this range \rangeh , it is not possible
to set $\tilde{N}=q=0$, and the lowest energy state 
will be in a continuous representation. 
In our semi-classical discussion in the last section, 
we found that the discrete representation can decay into
the continuous representation at $h = kw^2/4$. Now we see that, 
in the fully quantum description, the range over which a continuous
representation has lower energy has expanded from 
the point $h=kw^2/4$ to a strip of width $w$:
\eqn{strip}
{ {k \over 4} w^2 - {w \over 2} < h-1+ {1 \over 4(k-2)}
  < {k \over 4} w^2 + {w \over 2}.  }

So far we have restricted our attention to right-moving sectors
of the Hilbert space. Let us now discuss how the left and 
right movers are combined together. For the classical
solution of the long string, the worldsheet periodicity requires
that the spectral flow has to be done simultaneously
on both the left and right movers with the same amount. 
If $AdS_3$ were not the universal cover of $SL(2,R)$
but its single cover, different amounts of the left and
the right spectral flows would have been allowed since
the resulting solution is periodic modulo the closed 
timelike curve of $SL(2,R)$. 
It is straightforward to identify the corresponding constraint
in the quantum theory. 
Suppose we perform the spectral flows by the amount
$w_L$ and $w_R$ on the left and the right-movers. 
A state with conformal weights $(h_L, h_R)$ and
the $J_0^3$ charge $(\tilde{m}_L, \tilde{m}_R)$ is mapped by this
transformation to a state with conformal weights
$(h_L - w_L \tilde{m}_L - {k \over 4} w_L^2, 
h_R - w_R \tilde{m}_W - {k \over 4} w_R^2)$,
according to (\ref{virflow}).  
The worldsheet locality, which is the quantum counterpart
of the periodicity of the classical solution, requires
that the conformal weights $h_L$ and $h_R$ differ only 
by an integer. If this is the case before  spectral
flow, the same requirement after the flow implies
\eqn{flowconstraint}
{ w_L \tilde{m}_L + {k \over 4} w_L^2 = w_R 
\tilde{m}_R + {k \over 4}
    w_R^2 ~~~({\rm mod~integer}). }
For generic values of $(\tilde{m}_L, \tilde{m}_R)$, 
the only solution to this constraint is $w_L = w_R$.
In this paper, we are considering only the universal
cover of $SL(2,R)$ as the target space of the model.
In this case, the spectrum of $(\tilde{m}_L, \tilde{m}_R)$
is continuous, and only the left-right symmetric spectral
flow $w_L = w_R$ is allowed. 

\bigskip

\noindent
{\bf Summary:}
   
We propose that the spectrum of the $SL(2,R)$ WZW model 
(for the universal cover of $SL(2,R)$) contains 
the following two types of 
representations. First the spectral flow of the continous representations, 
with the same amount of spectral flow on the left and right, 
 $ \hat {\cal C}^{\alpha ,w}_{1/2 + is,L} 
\times \hat{\cal C}^{\alpha ,w}_{1/2 +is,R } $.
 Then the discrete representations
$\hat {\cal D}^{+,w}_{\tilde j,L} \times \hat {\cal D}^{+,w}_{\tilde j,R}$
with the same amount of spectral flow on the left and right and the same
value of $\tilde j$, with $ 1/2 < \tilde j < (k-1)/2 $. 
In the string theory, 
these representations should be tensored with the states of the internal
CFT, and the Virasoro constraints should be imposed.

\section{Scattering of long string}

When a long string comes in from the boundary of $AdS_3$ 
to the center, it will scatter back to the boundary. 
In this process the winding number could in principle change.
In order to study the S-matrix between incoming and outgoing
long strings, it is convenient to perform the rotations to 
Euclidean signature spaces, both on the worldsheet and in spacetime.
Following the standard procedure, we define the hermiticity as is natural in 
the Lorentzian theory. For this reason we still have 
the $SL(2,R)_L \times SL(2,R)_R$  currents in the Euclidean theory. 
The relevant conformal field theory, whose target space is
the 3-dimensional hyperbolic space $H_3 = SL(2,C)/SU(2)$ has
been studied in \cite{Gawedzki:1991yu,Teschner:1997ft,
Teschner:1999ug,Giveon:1998ns,deBoer:1998pp,Kutasov:1999xu,Giribet:1999ft}.

\subsection{Vertex operators}

To compute the scattering amplitudes, we would like to find vertex 
operators for all representations considered above. 
Spectral flow is realized in the vertex operator formalism in the following
standard fashion \cite{Gepner:1987hr}. 
We bosonize the $J^3$ currents,
introducing left and right moving chiral bosons\footnote{
Reflecting the hermiticity of the $SL(2,R)$ model, the scalar field $\phi$ 
is hermitian, but with a wrong sign for the
two-point function  $ \langle \phi(z) \phi(z') \rangle =
 \log(z-z')$.} through
\eqn{bosoniz}{
J^3_R = -i\sqrt{ k \over 2} \partial \phi(z) ~~~~~~~~~~
J^3_L = -i \sqrt{k \over 2} 
\bar \partial \phi(\bar z)
}
A state with charge $m$ under $J_R^3$ contains an exponential in $\phi(z)$
of the form $e^{ im \sqrt{ 2 \over k} \phi(z) } $.
The other two currents therefore can be expressed as 
\eqn{curr}{
J_R^+ = \psi e^{i\sqrt{ 2 \over k} \phi(z) },~~~ J_R^- = \psi^\dagger 
e^{-i\sqrt{ 2 \over k} \phi(z), }
}
and similarly for $J_L^\pm$. A primary field $\Phi_{jm \bar m}
(z, \bar z)$ of the current algebra can be expressed as
\eqn{decomp}{
\Phi_{j m \bar m} = e^{ i m \sqrt{ 2 \over k} \phi(z) + i\bar m
\sqrt{ 2 \over k} \phi(\bar z)} \Psi_{jm\bar m},
}
where $\Psi_{jm\bar m}$ carries no charges with
respect to $J_{R,L}^3$. In the case of the $SU(2)$ model,
the field corresponding to $\Psi$ is known as a parafermion. 
The parafermion for the $SL(2,R)$ model was studied in \cite{Lykken:1989ut}. 
The conformal weights of the parafermion field $\Psi_{jm\bar m}$
is 
\eqn{parafermionweight}
{ \eqalign{& h_{\Psi;jm\bar{m}}= - {j(j-1) \over k-2} + {m^2 \over k} \cr  
&\bar{h}_{\Psi;jm\bar{m}}= - {j(j-1) \over k-2} + {\bar{m}^2 \over k}. 
 }}
In the discrete lowest weight representation, $m, \bar{m} = j, j+1, j+2,
\cdots$. In particular, when $j=k/2$, the field $\Psi_{j= k/2,m=\bar{m} =k/2}$
has conformal weights $h = \bar{h} = 0$. 
Since the parafermion field lives in the unitary conformal field theory
it is natural to assume that it is the identity operator\footnote{Recently
we have learned that a similar argument has appeared in unpublished
notes by A. B. Zamolodchikov. We thank him for having his note available
to us \cite{Zamolodchikov:unpub}.}.  Here we simply note that the operator 
$$e^{i\sqrt{{k\over2 }}(\phi(z) + \phi(\bar{z}))} $$
has the correct OPE for the primary field of spin $j=k/2$
with the $SL(2,R)$ currents.

Using the parafermion notation, the operator obtained by
the spectral flow by $w$ units is expressed as
\eqn{decflw}{
\Phi^w =  e^{ i (\tilde m +w k/2)
 \sqrt{ 2 \over k} \phi(z) +i (\tilde{\bar  m }+ w k/2)
\sqrt{ 2 \over k}
\phi(\bar z)} \Psi_{j\tilde m \tilde{\bar m}}
}
It is easy to see that the conformal weight is given by 
\eqn{confwght}{
L_0 = { - j(j-1) \over k-2} - m w + k w^2/2
}

\subsection{Reflection coefficient}

We will compute the amplitude, using the formulae obtained
in 
\cite{Dorn:1994xn,Zamolodchikov:1996aa,Teschner:1997ft,Zamolodchikov:unpub},
in the case that the winding number
does not change.

The long string states are in the spectral flow of the continuum
representation. 
The corresponding vertex operators are 
\eqn{vercon}{ \eqalign{
\Phi^j_{ m {\bar m} } =&  e^{  m \phi(z) + {\bar m} 
\phi(\bar z)} \Psi^j_{ \tilde m \tilde{\bar m}} V_{h\bar h}(z,\bar z)
~,\cr
\tilde m = m - w k/2 ~,&~~~~~~\tilde {\bar m} = \bar m - w
k/2~,~~~~~
j={1\over 2} + i s}
}
where $V_{h \bar h}$ is an operator in the internal part with
conformal weights $(h,\bar h)$. 
The physical energy $E$ and angular momentum $\ell$ 
of a state in $AdS_3$ are given by
\eqn{physen}{
 m = {1 \over 2}(E +\ell) ,~~~~~~{\bar m} = {1\over 2}
(E -\ell).
}
The physical state constraint is \onshcon\ with $\tilde{N}=0$.
%
This 
 implies that
\eqn{ener}{
\tilde m = - w k/4 + {1 \over w} \left[
{ 1/4 + s^2 \over k-2} + h -1 \right]
}
Now we can now consider the  two point function 
\cite{Teschner:1997ft,Teschner:1999ug,Zamolodchikov:unpub} 
\eqn{zamtwo}{ 
\eqalign{ 
 \langle \Phi^j_{m \bar m}(z,\bar z) 
\Phi^{j'}_{ m' \bar m'}(z', \bar z')
 \rangle = &
{ \Gamma(1/2  +is - \tilde m) \Gamma( 1/2 +is + \tilde{\bar m}) 
\Gamma(-2 is) \Gamma( { 2 is \over 
k-2}) \over 
 \Gamma( 1/2 - is - \tilde{m}) \Gamma( 1/2 - is + \tilde{\bar m}) 
\Gamma(2 is) \Gamma(- { 2 is \over 
k-2})} \cr
&~~~\times \delta(s -s')\delta_{ N+N'} \delta (E + E') }}
The $z$ dependence is just $1/|z -z'|^4$ coming from the fact that the
two operators have weight (1,1). 
This is the reflection amplitude and the values of $\tilde m, \tilde{\bar m}$ are
determined
by \ener 
(notice that  $ m$ is the physical energy, not $\tilde m$.).

As explained in \cite{Giveon:1998ns} in this context, 
in string theory we have to integrate over $z$ and divide by the 
volume of $SL(2,C)$.  
We can use $SL(2,C)$ invariance to put $z=0$, $z' = \infty$ in the  
correlator. The volume of the rest of $SL(2,C)$ then gives
 $ \int { d^2 z \over |z|^2}$, which
cancels  one of the delta-functions in \zamtwo .
Notice that $\delta(s-s') \delta(E+E') =\delta(s-s') \delta(0)$,
the volume of $SL(2,C)$ 
cancels the $\delta(0) $ piece. 

Now if we study the poles of \zamtwo , we find that they are located 
at $1/2 +is - \tilde m = -q$ with $q=0,1,2, \cdots$. They
come from the first Gamma-function. Taking this condition
together
with \ener\ we find that 
\eqn{pole}{
1/2 +is +q = \tilde m = 
- w k/4 + {1 \over w} \left[
{ 1/4 + s^2 \over k-2} + h -1 \right]
}
and this equation is precisely the same as the usual mass shell equation
for discrete states if we take $\tilde j  = 1/2 + is$.
There are similar poles from the second Gamma-function.
There are no poles coming from the third factor since they cancel
extra poles appearing in the other factors. 
Notice that the poles appearing in \pole\ satisfy precisely the equation
\spectfl\ for bound states in the 
representation $\widehat {\cal D}^{+,w}_{\tilde j}$ (with $\tilde{N}=0$).
There is however an important difference. In \spectfl\ the value of 
$\tilde j$ obeyed the condition 
\eqn{rangej}{ 
{1 \over 2} < \tilde j < {k-1 \over 2}
}
 while we do not  have
such a condition in \pole . It is interesting to note that if $\tilde j$
satisfies \rangej\ then the residue at the pole has the proper sign to
be interpreted as coming from a bound state. When $\tilde j = (k-1)/2$,
$i.e$ at the upper bound of \rangej , we find that there 
is no pole. Moreover, immediately
above that value, we have the wrong sign for the pole residue. 
This might make us worry that the amplitude is not having the right
analytic structure. However, in order to have a one-to-one correspondence
between poles of the scattering amplitude and bound states, 
the potential has to decrease sufficiently rapidly at the infinity
\cite{landau}, a condition that is not met in our case. In such 
a situation, it is possible to have extra poles that do not correspond 
to physical states. 
We plan to analyze the poles and their implications for physical states
in a future publication.

\subsection{Relation to the scattering off the two-dimensional black hole}

The coset of the $SL(2,R)$ WZW by the $U(1)$ generated by $J^3$ gives 
a sigma-model whose target space is
the two-dimensional black hole with the Euclidean
signature metric \cite{Witten:1991yr}. The geometry of the black hole
is like a semi-infinite cigar with an asymptotic region in the
form of the cylinder $\IR \times S^1$. The dilaton field grows as
one approaches the center of the black hole, but it remains finite
since the geometry is terminated at the tip of the cigar. 
The string theory on $SL(2,R)/U(1) \times$(time)$\times {\cal M}$
is closely related to  
the string theory on $AdS_3 \times {\cal M}$ since the physical 
state conditions for the latter implies $J_{n}^3 
|{\rm physical}\rangle = 0$ for $n \geq 1$,
as we show in Appendix A. 
Similarly the superstring theory on $AdS_3 \times {\cal M}$
is related to the Kazama-Suzuki coset $SL(2,R)/U(1)$.

There is however difference between the zero mode sectors
of the theories on $AdS_3$ and on the two-dimensional black hole.  
In order to construct representations for $SL(2)/U(1)$, we can 
start from the representations of $\widehat{SL}(2,R)$ that we described
above and  impose the condition that $J^3_{n>0}$ annihilate the state and
that  the total $AdS_3$ energy 
vanishes, $J^3_{0,R} + J^3_{0,L} = m + \bar m =0$. 
In terms of the parafermion $\Psi_{j\tilde m \tilde{\bar m}}$ given in 
\decomp\ and \decflw , the condition is
$ \tilde m + \tilde{\bar m} = w k$. The locality condition
$  m -{\bar m} = n$ where $n$ is an integer implies 
that $\tilde m - \tilde{\bar m} = n$.
 These two quantization conditions are the ones
in \cite{Dijkgraaf:1992ba}  (see equation 3.6 of that paper).
The $SL(2)/U(1)$ theory has been studied recently in connection
with ``little'' string theories in \cite{Giveon:1999px,Giveon:1999tq}.

\section{Conclusion}

In this paper, we studied the physical spectrum of bosonic 
string theory in $AdS_3$. We proposed that the complete
Hilbert space of the $SL(2,R)$ WZW model consists of 
the continuous representations 
and their spectral flow 
$\widehat{\cal C}_{j=1/2 + is}^{\alpha,w}
\times \widehat{\cal C}_{j=1/2 + is}^{\alpha,w}$,
and the discrete representations and their spectral
flow $\widehat{\cal D}_j^{\pm,w}
\times \widehat{\cal D}_j^{\pm,w}$
with the constraint $1/2 < j < (k-1)/2$.
The sum over the spectral
flow is required if we assume that the Hilbert space
realizes the full loop group of $SL(2,R)$, including
its topologically non-trivial elements. We found
that this proposal leads to the physical spectrum
of the string theory with the correct semi-classical limits. 

In particular, we have solved the two puzzles which we mentioned
in the introduction. The no-ghost theorem for $\widehat{\cal D}_j^\pm$
requires the constraint $0< j < k/2$. If we only had the unflowed sector
(with $w=0$), it would imply the upper bound on 
allowed mass of string states, which appears artificial. 
This was one of the puzzles. We have resolved this puzzle
by showing
that the upper bound on the mass is removed if we include
all the spectral flowed sectors in the Hilbert space. 
Moreover we showed that the consistency with the spectral flow
and the standard harmonic analysis of the zero modes
requires the constraint $1/2< j < (k-1)/2$, more stringent than
the one required by the no-ghost theorem. 
The constraint $1/2 < j < (k-1)/2$ is found to be consistent 
with the locations of the poles in the reflection coefficient 
(with the correct sign for the pole residues; see also
\cite{Giveon:1999px})
 and 
the modular invariance of the partition function. 

Another puzzle was to identify states in the Hilbert space
corresponding to the long strings. We found that
these states are in the spectral flow of the continuous
representations, $\widehat{\cal C}_{j=1/2 + is}^{\alpha, w}
\times \widehat{\cal C}_{j=1/2 + is}^{\alpha, w}$. 
The integer $w$, which parametrized the amount of the spectral flow,
is identified with the winding number 
of the long string stretched closed the boundary of $AdS_3$. 
The physical spectrum of the long strings obtained from these 
representations agrees with the expectations from the semi-classical 
analysis in \cite{Maldacena:1998uz,Seiberg:1999xz}.

The resolutions of these puzzles removes 
the longstanding doubts about the consistency of the model. 
Moreover it appears that the $SL(2,R)$ WZW model is exactly 
solvable, just as WZW models for compact groups, although its 
Hilbert space structure is significantly different from those
of the compact cases. We hope that further study of the model 
will provide us more useful insignts into the $AdS$/CFT 
correspondence and strings in curved spaces in general.   

\section*{Acknowledgements}

We would like to thank A. Zamolodchikov for discussions 
and for giving us a copy of his unpublished notes. 
We also thank N. Seiberg, C. Vafa and  E. Witten 
for discussions. We would like to thank S. Hwang
for useful comments on the earlier version of this paper.

H.O.\ would like to thank J. Schwarz and 
the theory group at Caltech for the kind hospitality
while the bulk of this work was carried out. 
H.O.\ also thanks the hospitality of the theory group 
at Harvard University, where this work was initiated,
ICTP, Trieste and ITP, Santa Barbara, where
parts of this work were done. 

The research of J.M.\ 
was supported in part by DOE grant DE-FGO2-91ER40654,
 NSF grant PHY-9513835, the Sloan Foundation and the 
 David and Lucile Packard Foundations. 
The research of H.O.\ was supported in part by 
NSF grant PHY-95-14797, DOE grant DE-AC03-76SF00098,
and the Caltech Discovery Fund. 

\newpage

\appendix

\section{No-ghost theorems}

In this appendix we would like to extend the proof of the no-ghost theorem
to all the representations considered above. We assume $k>2$.
The proof of the no-ghost theorem for the standard lowest energy
representations \cite{Balog:1989jb,Petropoulos:1990fc,Mohammedi:1990dp,%
Bars:1991rb,Hwang:1991aq,Hwang:1992an,Hwang:1998tr,Evans:1998qu,Dixon:1989cg}
 involves two parts. 
Part I consists of  showing
 that a physical state can be chosen, up to a null state,
to be such that $J^3_n |\psi> =0$, for $n \geq 1$. This first part 
uses $0<j<k/2$ for the ${\cal D}^{\pm}_j$ representations as
well as $c=26$ and the mass shell condition. This was shown in 
\cite{Balog:1989jb,Petropoulos:1990fc,Mohammedi:1990dp,Bars:1991rb,%
Hwang:1991aq,Hwang:1992an,Hwang:1998tr,Evans:1998qu}.
Part II consists in showing that any state that is annihilated by
$J^3_{n>0}$ has non negative norm.  This step also uses $0<j<k/2$ for
the ${\cal D}^\pm_j$ representations. This was done in \cite{Dixon:1989cg}.
Here we will use the same strategy and prove Part I for the 
all our representations. The no-ghost theorem then follows from Part II.

\medskip

We first review the proof of Part I for the representations with $w=0$ and
then we do Part I for the $w \not =0$ representations.

\subsection{Proof of Part I for unflowed representations}

Here we follow the proof in \cite{Balog:1989jb,Petropoulos:1990fc,%
Hwang:1991aq,Hwang:1992an,Evans:1998qu}.
It has essentially three steps.

\medskip

\noindent
Step 1: The first step of the proof is to show 
 that states of the form
\eqn{basis}
{\eqalign{&L_{-n_1} L_{-n_2} \cdots L_{-n_N} J_{-m_1}^3 
J_{-m_2}^3 \cdots J_{-m_M}^3 |f\rangle \cr
&~~~~n_1 \geq n_2 \geq \cdots \geq n_N, ~~~
m_1 \geq m_2 \geq \cdots \geq m_M, \cr
 & {\rm with }~ L_n |f \rangle = J^3_n |f \rangle =0 ~~~{\rm for} 
~~n\geq 1,}} 
are linearly independent 
 and that they form a complete basis of the Hilbert
space. 

The states
 $|f\rangle$ are constructed from states in the current algebra times
some states in an internal conformal field theory. This internal piece is
assumed to be unitary.
This step involved separating  the piece of $L_n$ involving $L^{(3)}
= :J^3J^3:$, defining $ \hat L_n = L_n - L^{(3)}_n $.
One can show that the states \basis\ are in one to one correspondence
with states of the form 
\eqn{basisnew}{\eqalign{&L_{-n_1} L_{-n_2} \cdots L_{-n_N} J_{-m_1}^3 
J_{-m_2}^3 \cdots J_{-m_M}^3 |f\rangle \cr
&~~~~n_1 \geq n_2 \geq \cdots \geq n_N, ~~~
m_1 \geq m_2 \geq \cdots \geq m_M, 
}}
Notice that the conditions \basis\ on $|f\rangle$ are the 
same as $ \hat L_{n>0} |f \rangle = J^3_{n>0} |f \rangle =0$.
It is easier to show that \basisnew\ is a basis since now we can
think of the CFT as a product of a $U(1)$ factor with the rest.
The rest is a CFT with $c=25$ and therefore the fact that 
\basisnew\ is a basis reduces to showing that there are no 
null states in the Virasoro descendents on a primary field. 
This will be true if the conformal weight of the rest is positive.  
This  reduces to showing that
 $c_2/(k-2) + m^2/k +M >0$, where $M$ is the grade in the $SL(2,R)$ piece.
 For the continuous representations, this is
obvious since $c_2>0$. For lowest weight representations,
this inequality can be shown by rewriting it as
\eqn{twotwenty}{
{ 2 j \left(k/2 -j\right) \over k(k-2) } + { 2 M \over k}
\left( {k\over 2} -j\right) + { 2 j \over k}
( -j +m +M) + { 1 \over k} (j-m)^2 >0
}
We to use $0<j<k/2$ 
and also the fact that $m\geq j-M$, which is true
in general.
 Notice that the $m$ that
appears here is the total $J_0^3$ value, after we applied $J^\pm_{-n}$ 
any number of times. 
Notice that in this step we did not use that the states were obeying
the mass shell condition, but we used $0<j<k/2$ and that $c=26$.

\medskip
\noindent
Step 2: Here we show that a physical state can be chosen so that
it involves no $L_{-n}$ when written as \basis . 

 A physical state can be written as a state with no $L_{-n} $  plus
a spurious state. A spurious state is a state with at least one $L_{-n}$.
Then we use the fact that, when $c=26$,  $L_n$ ($n\geq 1$)  acting on
  a spurious state which
satisfies the $L_0=1$ condition leaves it as a spurious state
\cite{Brower:1972wj,Goddard:1972iy} .
If $L_{n>0}$ acts on a state of the form \basis\ with no $L_{-n}$ 
then it will not produce any $L_{-n}$. Together with
the fact that \basis\ is a basis this implies that 
the part of the state with no $L_{-n}$ satisfies the physical state
condition on its own, and therefore the rest is a null state (a spurious
physical state).

\medskip
\noindent
Step 3: We show that if the physical state $| \psi \rangle$
 involves no $L_{-n}$ when
written as in \basis\ then $J^3_n | \psi \rangle =0$. 

 Since there are no $L_{-n}$'s in the physical state $\psi$ this
implies that $L^{(3)}_n \psi =0$ for $n\geq 1$. Then we try to show that
the only states satisfying this will be states with $J^3_n\psi =0$ for
$n\geq 1$. This would be true if there are no null states in the $L^{(3)}$
Virasoro
descendents of the states $|f\rangle$ we considered above.
If $m \not =0$ then one can show that there is no null state in the 
Virasoro descendents in the $L^{(3)}$ Virasoro descendents.
There are two states with $m=0$ one is in the continuous representation, 
but the mass shell condition automatically implies that $N=0$ (there
are no $J^a_{-n}$ in this state) and therefore the state has positive norm.
The other is the state in the lowest weight representation
\eqn{zeroen}{
J^-_{-1} | j=1 \rangle} 
 (and of course its complex conjugate in the
highest weight representation). This state has positive norm.
Note that $m$ is the physical energy in $AdS_3$ of the
state
in question. Zero  energy states, therefore imply that we have
a normalizable 
 zero mode.
This is the state corresponding to the identity operator in the
spacetime boundary conformal field theory, the state $\bar J J \Phi_1$ of 
\cite{Kutasov:1999xu} which played an important role in the 
computation of the spacetime Virasoro algebra. 

One can show, using the mass shell condition, that all other states
have $m\not =0$. 
The mass shell condition is
\eqn{massshell}{
- {j(j-1) \over k-2} + N + h' -1 =0
}
where $N$ is the grade in the $SL(2,R)$ part and $h'$ is the conformal weight
of the rest, $h' \geq 0$. If $0< j < 1$ then $m$ is nonzero because
it can only change by an integer by the action of the $J^\pm_n$ currents.
If $j=1$ with   $N=1$ and $h'=0$ we find \zeroen  and states with positive
$m$. 

Consider now $j>1$. If we had  $m=0$ then we also need   $N\geq j$, $j \geq
2$ (since $m=0$ only if $j $ is integer) and
furthermore
\eqn{newineq}{
 - {j(j-1) \over k-2} 
+ N -1 \geq  {(j-1) ( k - 2 -j ) \over k-2} >0
}
provided $j
\leq k/2 $. Since $j$ has to be at least 2, then $k>4$ and therefore
 $k-2-k/2 >0$. Thus we conclude
 that \massshell\ would not be obeyed if $m=0$.

\subsection{Proof of Part I for flowed representations}

Now we would like to generalize the above discussion to 
the spectral flowed   representations that we called 
$\hat {\cal C}^{\alpha, w}_{1/2 + is} $ and $\hat {\cal D}^{+,w}_{\tilde j}$.
In the case of discrete representations we want to show that the
no ghost theorem holds for $0 < \tilde j < k/2$ where $\tilde j$ labels 
the representation before we perform the spectral flow operation, i.e.
it labels a representation of the current algebra with $\tilde L_0$
 bounded below. 
So we consider the same representations we had above but we
 modify the physical state conditions. 
This is equivalent to imposing the usual conditions on the flowed 
representations. 
 We would like to prove that, given any state built on a
lowest weight or continuous representation with respect to
$\tilde{J}_n$, the physical state condition $(L_n -\delta_{n,0})
|\psi\rangle =0$ $n\geq 0$  with respect to $L_n$
removes non-negative norm states.
We only consider spectral flow with $w>1$ on continuous or 
lowest weight representations $\hat {\cal D}^+_{\tilde j}$. These and their 
complex conjugates cover all the representations we needed to consider.
We reproduce now the  steps in A.1.

\medskip
\noindent
Step 1:
In (\ref{basis}) we need to show that they form a basis with
 $L_{-n} =  \tilde L_{-n} 
- w \tilde J_{-n}^3$. We know that they would form a basis if we had
an expression like \basis with $L_{-n} \to \tilde L_{-n}$. Fortunately
there
is an invertible one to one map between these two sets of states, so that
they form a basis.

\medskip
\noindent
Step 2: It is the same since only $c=26$ is used.

\medskip
\noindent
Step 3:
If we write a physical state, $|\psi\rangle$,
  as a state  with no $L_{-n}$ then
 $L^{(3)}_n  $ with $n\geq 1$
annihilates it. Again we will try to show that $m=
\tilde{m} + kw/2$ is
nonzero and that will imply that $J^3_{n>0} |\psi\rangle =0$.
For this we need to use the new mass shell condition
\eqn{newmass}{
{\tilde c_2 \over k-2} + \tilde{N} + h' - w  m + { k w^2 \over 4} =1
}
where $\tilde{N}$ is the level inside the current algebra before
the spectral flow, $\tilde c_2 $ is the second
casimir in terms of $\tilde j$  and $h'$ is the conformal
weight
of state in the internal  conformal theory 
(the internal piece needs not be a primary state,
and we only require that the whole combined state
needs to be primary).
We can assume with no loss of generality that $w\geq 1$.
Let us start with  the spectral flow of a continuous representation,  \newmass 
implies that if $m=0$ then $\tilde N=0$ 
and and there are no negative norm states.
(The only solution with $m=0$  is in the case of $k=3$ and $\tilde j=1/2$). 

Let us turn to lowest weight representations.
Thanks to the restriction $0<\tilde j < k/2 $,
 we have $\tilde c_2/(k-2) > -k/4$.
Therefore, if $m=0$, the left hand side of \newmass\ is 
larger than $ k/4( w^2 -1) $. If $w \geq 2$,
\newmass\ cannot be obeyed.
If $w=1$, $m=0$ implies $\tilde{m} = - k/2$ 
and 
$\tilde{N}$ in \newmass\ has to be at least $\tilde{N}\geq \tilde j +k/2$. 
However, in this case we find 
$\tilde c_2/(k-2) +\tilde{N} + k/4   \geq k/2 + \tilde j > 1$  (here we used $k>2$)
and again \newmass\ is not satisfied. 

So we conclude that all states can be mapped into states obeying
$J^3_{n>0} |\psi \rangle =0$.

\section{Partition function}

In this Appendix, we discuss the partition function of
the $SL(2,R)$ WZW model and its modular invariance.

\subsection{Partition function of the SU(2) model}

Before we begin discussing the modular invariance of
the $SL(2,R)$ theory, let us review the case of $SU(2)$. 

The characters $\chi_l^{k}(\tau,\theta)$ ($l=0, {1\over 2}, 1, \cdots
{k \over 2}$)
of the irreducible representations of the $SU(2)_k$ affine algebra
transform under the modular transformation as
\eqn{sumod}{
  \chi_l^k(-1/\tau, -\theta/\tau) = 
  {\rm exp} \left(2\pi i {k \over 4} {\theta^2 \over \tau} \right) 
  \sum_{l'} S_{ll'} \chi_{l'}^k(\tau,\theta),}
where $S_{ll'}$ is some orthonormal $(k+1) \times (k+1)$
matrix. The diagonal (so-called $A_k$-type) modular invariant
combination is therefore
\eqn{suinv}{
   e^{-2\pi {k \over 2} {(Im \theta)^2 \over Im \tau}}
     \sum_l |\chi_l(\tau,\theta)|^2.}
The exponential factor  
$e^{-2\pi {k \over 2} {(Im \theta)^2 \over Im \tau}}$ 
is there to cancel the exponential factor in \sumod\ as
\eqn{cancel}{{[Im (-\theta/\tau)]^2 \over Im (-1/\tau)}=
{(Im \theta)^2 \over Im \tau}+ i{\theta^2 \over 2 \tau}
- i {\bar{\theta}^2 \over 2\bar{\tau}}. } 
It is known that the exponential factor in \suinv\ is a consequence
of the chiral anomaly and therefore of the OPE singularity,
\eqn{suope}{ J^3(z) J^3(w) \sim {k/2 \over (z-w)^2} .} 

\subsection{Partition function of the SL(2,C)/SU(2) model}


In string theory, one-loop 
computations are done after performing the
Euclidean rotation on both the target space
and the worldsheet (or stay in the Lorentzian 
signature space and use the $i\epsilon$ prescription). 
The modular invariance of the partition function
is imposed on the Euclidean worldsheet. 
In our case, the Euclidean rotation of the target space
means $SL(2,R) \rightarrow H_3=SL(2,C)/SU(2)$. 
The partition
function of the $SL(2,C)/SU(2)$ model
has been evaluated in \cite{Gawedzki:1991yu}
as
\eqn{gawedzkipartition}{ Z_{SL(2,C)/SU(2)} 
 \sim {1\over \sqrt{Im \tau} e^{-2\pi (Im \theta)^2
\over Im \tau} |\vartheta_1(\tau,\theta)|^2}.}
Note that our 
definition of the partition function differs
from that in \cite{Gawedzki:1991yu} by the
factor $e^{2\pi {k \over 2} {(Im \theta)^2 \over
Im \tau}}$. It apears that, without this factor,
the partition function is not modular invariant\footnote{The
puzzle about the apparent lack of the the modular invariance 
was recently resolved in \cite{mos}.}.
One may expect
that this partition function is related to the
one for the $SL(2,R)$ model by the Euclidean rotation. 
In the discussion below, we first evaluate
the $SL(2,R)$ partition function on the Lorentzian
torus, and therefore take $\tau, \bar \tau, \theta,
\bar \theta$ to be independent real variables.
We then analytically continue them to complex values so
that $(\tau,\theta)$ are complex conjugate of $(\bar\tau,
\bar\theta)$. 
We will find that, by doing this analytic continuation, 
and ignoring  contact terms,
the $SL(2,R)$ partition function turns into
the $SL(2,C)/SU(2)$ partition function \gawedzkipartition , 
provided we impose the constraint
$1/2 < j < (k-1)/2$ on the discrete representations.

\subsection{Discrete representations of SL(2,R)}

The character of the discrete representation
$D_j^+$ is
\eqn{ch}{
\eqalign{
  \chi^+_j(\tau,\theta) = & 
 {\sl Tr}(e^{2\pi i \tau (L_0 - {k \over 8(k-2)})}
           e^{2\pi i \theta J_0^3}) \cr
= & { {\rm exp} \left[ 2\pi i \tau  \left(
   -{j(j-1) \over k-2} - {k \over 8(k-2)} \right)
  + 2\pi i \theta j \right] \over 
    (1-e^{2\pi i \theta}) \prod_{n=1}^\infty
      (1-e^{2\pi i n \tau}) 
 (1-e^{2\pi i n \tau}e^{2\pi i \theta}) 
 (1-e^{2\pi i n \tau}e^{-2\pi i \theta}) } \cr
= & { {\rm exp} \left[ - {2\pi i \tau \over k - 2} 
   (j- {1\over 2})^2 
+ 2\pi i\theta (j - {1 \over 2}) \right] \over 
  i \vartheta_1(\tau,\theta)}
    . }}
where $\vartheta_1(\tau,\theta)$ is the elliptic theta-function
\eqn{thetafun}{
 \vartheta_1 (\tau,\theta) = -i \sum_{n=-\infty}^\infty
  (-1)^n {\rm exp}\left[ \pi i \tau\left(n-{1\over 2}\right)^2 + 2\pi i \theta
             \left(n - {1 \over 2}\right)\right]. }
The spectral flow 
\eqn{spectral}{ \tilde{L}_0 = L_0 + w J_0^3 - {k \over 4} w^2,
~~ \tilde{J}_0^3 = J_0^3 - {k \over 2} w, ~~~
(w = 0 , \pm 1, \pm 2, \cdots), }
transforms the character $\chi_j^+$ as
\eqn{characterflow}{
\eqalign{
& {\sl Tr}(e^{2\pi i \tau (\tilde{L}_0 - {k \over 8(k-2)})}
           e^{2\pi i \theta \tilde{J}_0^3}) \cr
= &
  {\sl Tr}(e^{2\pi i \tau (L_0  + w J_0^3 - {k \over 4} w^2
 - {k \over 8(k-2)})}
           e^{2\pi i \theta (J_0^3- {k \over 2} w)}) \cr
= & 
{ {\rm exp} \left[ - 2\pi i \tau\left( 
   {(j- {1\over 2})^2 \over k - 2}  - w (j - {1\over 2}) 
      + {k \over 4} w^2  \right)
+ 2\pi i \theta (j - {1 \over 2} - {k \over 2} w) \right] \over 
  i \vartheta_1(\tau,\theta+ w \tau)} \cr
= & (-1)^w
{ {\rm exp} \left[ - {2\pi i \tau \over k - 2} 
   (j- {1\over 2} - {k-2 \over 2} w)^2
+ 2\pi i\theta (j - {1 \over 2} - {k-2 \over 2} w) \right] \over 
  i \vartheta_1(\tau,\theta)}  
, }}
where we used
\eqn{thetashift}{
 \vartheta_1(\tau, \theta + w \tau ) = (-1)^w {\rm exp}\left(
 - \pi i \tau w^2 - 2\pi i \theta w \right) \vartheta_1(\tau,\theta).} 
We have also performed
an analytic continuation such as
$$ \sum_{n=0}^\infty q^n = - \sum_{n=1}^\infty q^{-n},$$ 
ignoring terms like $\sum_{n=-\infty}^{\infty} q^n \sim
\delta(\tau)$. From here on, we allow $(\tau, \theta)$
to take complex values and $(\bar{\tau}, \bar{\theta})$
to be their complex conjugates. 
 
Let us sum over allowed representation.
According to our proposal about the Hilbert space
of the WZW model, all the representations in the allowed 
range  $1/2 < j < (k-1)/2$ should appear. 
We also require that the spectrum to
be invariant under the spectral flow \spectral , so we need
to sum over $w$. The part of the partition function 
made by discrete representations is then
\eqn{partition}{
\eqalign{
& e^{+ 2\pi {k \over 2} {(Im \theta)^2 \over Im \tau}}
   \sum_{w=-\infty}^\infty 
  \int_{1/2}^{(k-1)/2} dj 
 {{\rm exp} \left[ {4\pi  Im \tau \over k - 2} 
   (j - {1\over 2} - {k-2 \over 2} w)^2
- 4 \pi  Im \theta (j - {1 \over 2} - {k-2 \over 2} w) \right] 
 \over   | \vartheta_1(\tau,\theta)|^2}\cr
& = e^{+2\pi {k \over 2} {(Im \theta)^2 \over Im \tau}}
   \int_{-\infty}^\infty dt 
 {{\rm exp} \left[ {4\pi  Im \tau \over k - 2} 
   t^2
- 4 \pi  Im \theta t \right] 
 \over   | \vartheta_1(\tau,\theta)|^2}\cr
& \sim {1
       \over \sqrt{ Im\tau}
 e^{ -2\pi { ( Im\theta)^2 \over  Im\tau}}|\vartheta_1(\tau,\theta)|^2}
 . } }
It is interesting to note that the $j$-integral over the
range $1/2 < j < (k-1)/2$ and the sum over $w$ fit together to
give the $t$-integral over $-\infty < t < \infty$.  
Since the spectral flow with $w=1$ maps $D_j^+$ to $D_{k/2-j}^-$,
we do not have to consider the orbit of $D^-_j$ separately. 
The exponential factor $e^{+ 2\pi {k \over 2} {(Im \theta)^2 \over Im
\tau}}$
is due to the chiral anomaly, as in the $SU(2)$ case. The sign
in the exponent is opposite here since the sign
of the OPE of $J^3$ is opposite in the $SL(2,R)$ case.

The partition function computed in \partition\ is manifestly
modular invariant. In fact, it is identical to (\ref{gawedzkipartition})
computed for the $SL(2,C)/SU(2)$ model. This gives an additional
support for our claim that the Hilbert space of the $SL(2,R)$
model contains the discrete representations of $1/2 < j
< (k-1)/2$ and their spectral flow. 

The construction of the partition function here is
closely related to the one given in \cite{Henningson:1991jc}.
There, instead of the integral over $j$ in \partition ,
the partition function was given by a sum over integral
values of $j$. This is because they considered the string theory
on the single cover of the $SL(2,R)$ group manifold with the closed 
timelike curve. The resulting partition function, after analytic
continuation, is also modular invariant and appears to be a correct
one for such a model. It is, however, different from the partition function
(\ref{gawedzkipartition}) of the $SL(2,C)/SU(2)$ model, as it should  
since the Euclidean rotation of the $SL(2,C)/SU(2)$ model is naturally
related to the model on the universal cover of $SL(2,R)$ rather
than on its single cover.

\subsection{Continuous representations}

It is curious that the sum over the discrete representations
and their spectral flow alone reproduces the partition
function of the $SL(2,C)/SU(2)$ model. In fact, the sum over
the continuous representations and their spectral flow, although
formally modular invariant by itself, does
not contribute to the partition function if we assume
the analytic continuation in $\tau, \bar \tau, \theta, \bar\theta$ 
and ignore contact terms. 

The character of the continuous representation is parametrized
by a pair of real numbers $(s, \alpha)$ with $0 \leq \alpha < 1$
and $s $ arbitrary. The character is given by
\eqn{contch}{
\chi_{j=1/2 +is, \alpha} = \eta^{-3} e^{2\pi i {s^2 \over k - 2}\tau}
   e^{i\alpha \theta } \sum_n e^{2\pi i n \theta}. }
As before, we regard the worldsheet metric to be of
the Minkowski signature, and $\theta$ is real. So the
sum $\sum_n$ in the definition of $\chi_{j, \alpha}$
gives the periodic delta-function 
\eqn{sumdelta}{
 \sum_n e^{2\pi i n\theta} = 2 \pi \sum_m \delta(\theta + m).}
After the spectral flow \spectral, the character becomes
\eqn{contspec}{
\chi_{j=1/2+is,\alpha; w} = \eta^{-3} e^{2\pi i \left({s^2 \over k - 2}
 + {k \over 4} w^2 \right)\tau}
   2 \pi \sum_m e^{2\pi i m (\alpha - {k \over 2} w)}
 \delta(\theta + w \tau + m). }

Now let us take $|\chi_{1/2+is,\alpha ; w}|^2$ and integrate over
$s$ and $\alpha$. The integral over $\alpha$ forces $m_L = m_R$
in the summation in \contspec. The integral over $s$
gives the factor $1/\sqrt{Im \tau}$. So we have
\eqn{afterint}{
 \int_{-\infty}^\infty ds \int_0^1 d\alpha |\chi_{1/2+is,\alpha ; w}|^2
 = e^{-4\pi Im \tau {k \over 4} w^2}{1 \over \sqrt{Im \tau}
      |\eta|^6} \sum_{m} \delta^{(2)}(\theta + w \tau + m). }
Let us sum this over $w$. We get a non-zero result only when
there is some integer $w$ such that
\eqn{sol}{ w = - {Im \theta \over Im \tau}. }
Therefore
\eqn{modinvcont}{
 \eqalign{
& e^{+ 2\pi {k \over 2} {(Im \theta)^2 \over Im \tau}}
  \sum_w  \int_{-\infty}^\infty ds \int_0^1 d\alpha  |\chi_{1/2+is,\alpha  ;
w}|^2 \cr
& =  {1 \over \sqrt{Im \tau}
      |\eta|^6} \sum_{w, m} \delta^{(2)}(\theta + w \tau + m) .}
}
This expression is formally modular invariant since
$\sum_{w,m}$ sums over the modular orbit of the delta-function
and $1/|\eta|^4$ cancels its modular weight. If we assume the
analytic continuation, terms of this form are all set equal
to zero. So, in this sense, the continuous representation
does not contribute to the partition function of the
$SL(2,C)/SU(2)$ theory after the Euclidean rotation.

\renewcommand{\baselinestretch}{0.87}

\bibliography{paper}
\bibliographystyle{ssg}


\end{document}